\documentclass{article}
\usepackage{tabularx}
\usepackage{victor}
\usepackage{dsfont} 
\usepackage{cleveref}
\creflabelformat{equation}{#2\textup{#1}#3}
\usepackage{enumitem} 

\usepackage{pdflscape}   
\usepackage{array}       

\usepackage[sorting=none, 
			backend=biber,
            natbib=true,
			style=ext-authoryear,
			maxbibnames=999,
			maxcitenames=2,     
			mincitenames=1,     
			uniquelist=false,    
			giveninits=true,
			articlein=false,
			dashed=false,
			]{biblatex}
\AtEveryBibitem{
    \clearfield{urlyear}
    \clearfield{urlmonth}
	\clearfield{month}
	\clearfield{issn}
	\clearfield{isbn}
	\clearfield{editor}
	\clearfield{pages}
	\clearfield{volume}
	\clearfield{number}
	\clearfield{primaryclass}
}
\allowdisplaybreaks 

\title{Multi-scale species richness estimation with deep learning}
\author{Victor Boussange$^{1,\star}$, 
Bert Wuyts$^{1}$,
Philipp Brun$^{1}$, 
Johanna T. Malle$^{1,2}$, 
Gabriele Midolo$^{3}$, 
Jeanne Portier$^{4}$,
Théophile Sanchez$^{5,6}$,
Niklaus E. Zimmermann$^{1}$,
Irena Axmanová$^{7}$,
Helge Bruelheide$^{8,9}$,
Milan Chytrý$^{7}$,
Stephan Kambach$^{8}$,
Zdeňka Lososová$^{7}$,
Martin Večeřa$^{7}$,
Idoia Biurrun$^{10}$,
Klaus T. Ecker$^{11}$,
Jonathan Lenoir$^{12}$,
Jens-Christian Svenning$^{13}$,
Dirk Nikolaus Karger$^{1}$
\\
\begin{flushleft}
$^1$ \small{Dynamic Macroecology, Land Change Science, Swiss Federal Research Institute WSL, CH-8903 Birmensdorf, Switzerland}\smallskip\\
$^{2}$ \small{Department of Evolutionary Biology and Environmental Studies, University of Zurich, Zurich, Switzerland}\smallskip\\
$^{3}$ \small{Department of Spatial Sciences, Faculty of Environmental Sciences, Czech University of Life Sciences Prague, Praha - Suchdol, Czech Republic}\smallskip\\
$^4$ \small{Resource Analysis, Forest Resources and Management, Swiss Federal Research Institute WSL, CH-8903 Birmensdorf, Switzerland}\smallskip\\
$^{5}$ \small{Ecosystems and Landscape Evolution, Department of Environmental Systems Science, ETH Zürich, Zürich, Switzerland}\smallskip\\
$^{6}$ \small{Ecosystems and Landscape Evolution, Land Change Science, Swiss Federal Research Institute WSL, CH-8903 Birmensdorf, Switzerland}\smallskip\\
$^{7}$ \small{Department of Botany and Zoology, Faculty of Science, Masaryk University, Brno, Czech Republic}\smallskip\\
$^{8}$ \small{Institute of Biology, Geobotany and Botanical Garden, Martin Luther University Halle-Wittenberg, Halle (Saale), Germany}\smallskip\\
$^{9}$ \small{German Centre for Integrative Biodiversity Research (iDiv) Halle-Jena-Leipzig, Leipzig, Germany}\smallskip\\
$^{10}$ \small{Department of Plant Biology and Ecology, Faculty of Science and Technology, University of the Basque Country UPV/EHU, Bilbao, Spain}\smallskip\\
$^{11}$ \small{Ecosystem Dynamics, Biodiversity and Conservation Biology, Swiss Federal Research Institute WSL, CH-8903 Birmensdorf, Switzerland}\smallskip\\
$^{12}$ \small{UMR CNRS 7058 ‘Ecologie et Dynamique des Systèmes Anthropisés’ (EDYSAN), Université de Picardie Jules Verne, F-80037 Amiens, France}\smallskip\\
$^{13}$ \small{Center for Ecological Dynamics in a Novel Biosphere (ECONOVO), Aarhus University, DK-8000 Aarhus C, Denmark}\smallskip\\

$^\star$ Corresponding author, email: \href{mailto:vic.boussange@gmail.com (VB)}{\texttt{vic.boussange@gmail.com}}
\end{flushleft}
}
\date{\today}

\begin{document}

\maketitle

\begin{abstract}
Biodiversity assessments depend critically on the spatial scale at which species richness is measured. How species richness accumulates with sampling area is influenced by natural and anthropogenic processes whose effects vary across spatial scales. These accumulation dynamics, described by the species-area relationship (SAR), are challenging to assess because most biodiversity surveys cover sampling areas far smaller than the scales at which these processes operate.
Here, we combine sampling theory with deep learning to estimate species richness at arbitrary spatial scales across geographic space from existing ecological surveys.
We apply our model, named \modelname{}, to \totkEVAplots{} vegetation surveys across Europe. Validated against independent regional plant inventories, \modelname{} reduces root mean squared error of vascular plant richness estimates by 61\% relative to conventional estimators, yields substantially less biased predictions, and produces multi-scale richness maps alongside spatially explicit estimates of the species accumulation rate, a key indicator for biodiversity conservation.
By encompassing the full spectrum of ecologically relevant spatial scales within a single unified framework, \modelname{} provides an essential tool for robust biodiversity assessments and forecasts under global change.
\end{abstract}

\section{Introduction}
Biodiversity is undergoing unprecedented changes due to global change stressors \citep{ipbes_global_2019}, with impacts that critically vary across spatial scales \citep{smart2006}. Agriculture, for example, can increase local species richness by introducing ruderal and generalist species and by boosting productivity, yet simultaneously drives biodiversity loss at regional scales through habitat degradation and the homogenisation of species assemblages \citep{fahrig1997, maxwell2016, simkin2022, keil2015}. Similarly, climate change drives divergent richness patterns across spatial scales by altering species distributions and community assembly processes \citep{ipbes_global_2019}.
Assessing how diversity responds to global change across spatial scales is therefore crucial for evaluating the state and trends of biodiversity \citep{powers2019, newbold2018, he2011, rahbek2005}.
Yet, expert biodiversity surveys are labor-intensive endeavors that can only deliver sparse, small-scale biodiversity data. Upscaling these data to derive regional or global trends poses a major challenge, resulting in a critical knowledge gap \citep{ipbes2018}.

\begin{figure}[h]
    \centering
    \includegraphics[width=0.8\textwidth]{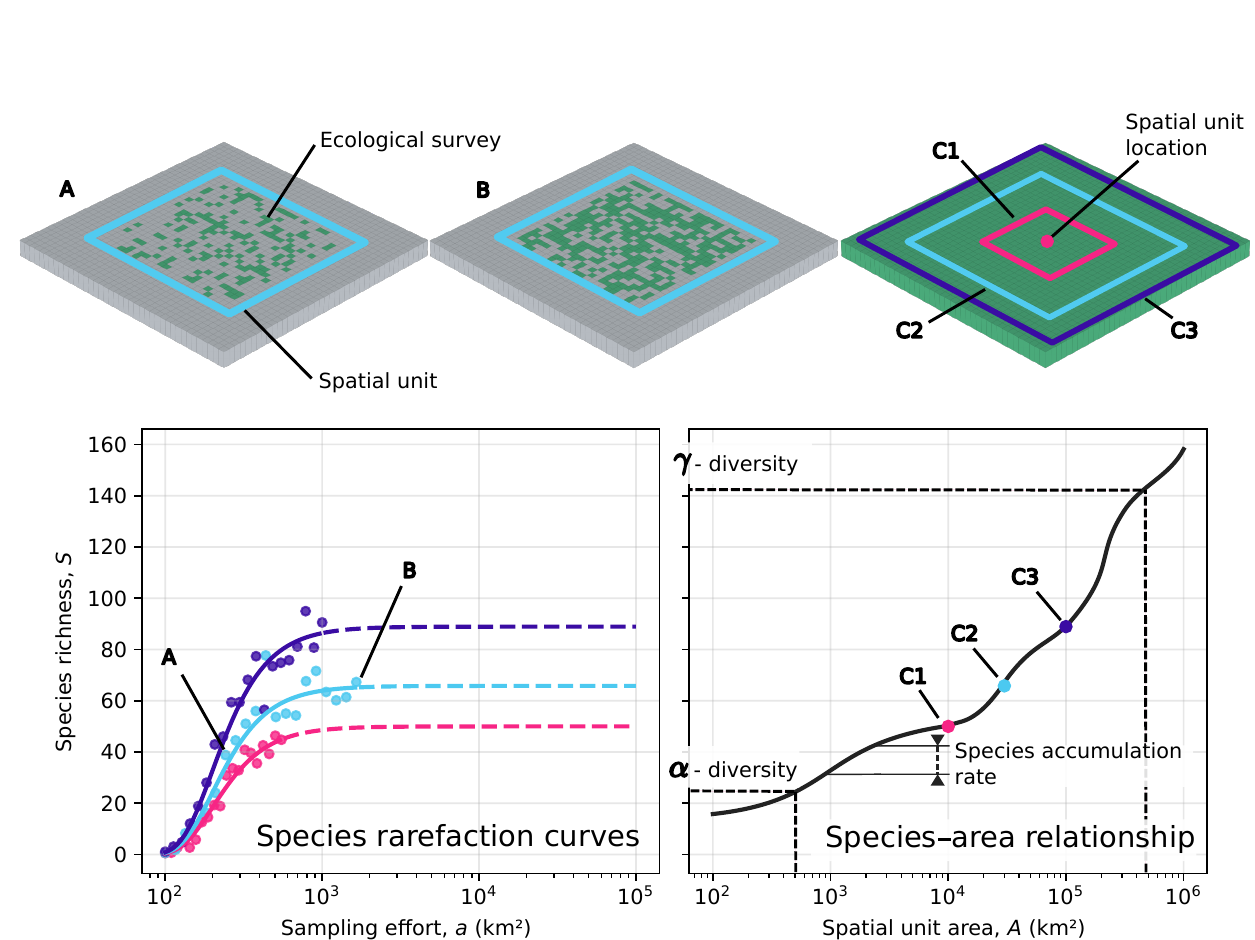}
    \caption{
        \textbf{Illustration of species rarefaction curves and species-area relationships.} Rarefaction curves (left plot) show how expected species richness $S$ increases with sampling effort $a$ within a spatial unit of fixed area $A$ and location. Red, blue, and purple rarefaction curves correspond to nested spatial units sharing the same centroid (location) but increasing areas, as illustrated in the top right panel. Rarefaction curves are estimated by fitting a parametric rarefaction model to ($a$, $S$) pairs obtained by randomly pooling biodiversity plots (as represented by green squares in the top panels). Each curve is bounded by total species richness $S_T$ in the unit, estimated by extrapolating the fitted rarefaction model under asymptotic sampling effort ($a \to \infty$, dashed lines). Continuously varying $A$ for a fixed centroid yields the species-area relationship (SAR; right plot), which relates $S_T$ to $A$.
        From the SAR, we can derive local ($\alpha$) and regional ($\gamma$) species richness, as well as the species accumulation rate (slope of the SAR), an indicator of species turnover (right plot).
    }
    \label{fig:src_sar}
\end{figure}

%
The species-area relationship (SAR) is a key concept to comprehend how biodiversity patterns change across spatial scales. Here, we use the term "spatial scale" broadly to designate the sampling area, study extent, spatial resolution, or spatial grain, \textit{sensu} \cite{o1986hierarchical}. Traditionally modelled as a power law, $S_T = cA^z$ \citep{arrhenius1921}, the SAR describes how total species richness $S_T$ increases with sampling area $A$. The parameter $c$ represents the species richness observed within a unit area, commonly referred to as $\alpha$-diversity. The parameter $z$ determines the slope of the SAR on a log-log scale (i.e., the derivative of $\log S_T$ w.r.t $\log A$), with higher $z$ values indicating a faster accumulation of species with area. $z$ is often associated with $\beta$-diversity or species turnover \citep{matthews2021a, halley2013, whittaker2007}, though these terms carry alternative definitions \citep{Tuomisto2010}; hereafter, we refer to the slope of the SAR as the species accumulation rate. Estimating the SAR at a certain location permits scaling up local $\alpha$-diversity to estimate regional species richness, also known as $\gamma$-diversity (\cref{fig:src_sar}). By allowing the prediction of species richness across the full range of ecologically relevant spatial scales, the SAR is instrumental for a comprehensive estimation of biodiversity.

While the power-law SAR model is a convenient approach to estimating species richness, it relies on the strong assumption of a linear relationship between area and species richness on a log-log scale, which may be considered only valid for a limited range of scenarios and spatial scales \citep{dengler2009, guilhaumon2008}. In reality, the accumulation of species richness changes in a non-linear fashion as different environmental conditions and dispersal barriers (e.g., mountain ranges, water bodies) are encountered at increasingly larger spatial extents \citep{conor2013, chisholm2025, borda-de-agua2025}. Biotic and abiotic drivers, such as climate, topography, soil conditions, and habitat type, among others, shape the spatial distribution of species, affecting the rate at which they accumulate \citep{he2002, he2011, weigelt2013, matthews2021a, chisholm2025, borda-de-agua2025, ulrich2022, drakare2006}. Importantly, the effect of these drivers is also scale dependent \citep{willis2002, wiens1989, turner2005, drakare2006}.

Several studies have characterized the effect of environmental drivers on the SAR \citep{turner2005}. One branch of research has drawn on exhaustive species inventories from islands or administrative units, such as countries or regional states, to investigate these effects. For instance, \citet{triantis2003, kalmar2007, kallimanis2008} showed that expressing the species accumulation rate as a function of habitat heterogeneity and climate better captures spatial variation in the SAR. However, the sparsity and coarse spatial scales of such political or geographical units limit the characterization of SAR variation at fine spatial grains.

To resolve species richness at finer spatial scales, another line of research has developed upscaling methods that estimate species richness from small-scale biodiversity plots (i.e., delimited sampling areas in which species occurrences are recorded, also known as quadrats or \textit{relevés}; \cite{kunin2018,portier2022,kusumoto2023}). These approaches use extrapolation methods to estimate total species richness within a spatial unit based on small-scale biodiversity plots \citep{chao2014,elenaschmitz2025}. One prominent method involves fitting a parametric function that relates species richness to sampling effort, known as the species rarefaction curve, which describes how species richness increases with sampling effort within a given spatial unit (see \cref{fig:src_sar} for an illustration). We define throughout the main text \textit{sampling effort} as the sum of the biodiversity plot areas used to calculate species richness (see \cref{sec:definitions}). Unlike the SAR, which typically relates richness to a single contiguous area, the rarefaction curve describes how expected richness grows with sampling effort from biodiversity plots scattered opportunistically or systematically within the considered spatial unit. The rarefaction curve is necessarily bounded by the total species richness within the spatial unit, which can be estimated by extrapolation as sampling effort approaches infinity \citep{soberonm.1993, gotelli2001}.
Conditioning this parametric function on the spatial unit's area and associated environmental conditions, which characterize each location in geographic space, could enable the prediction of location- and area-specific rarefaction curves. By extrapolation under asymptotic sampling effort, it could provide estimates of species richness at arbitrary locations and spatial grains.

\begin{figure}[h]
    \centering
    \includegraphics[width=\textwidth]{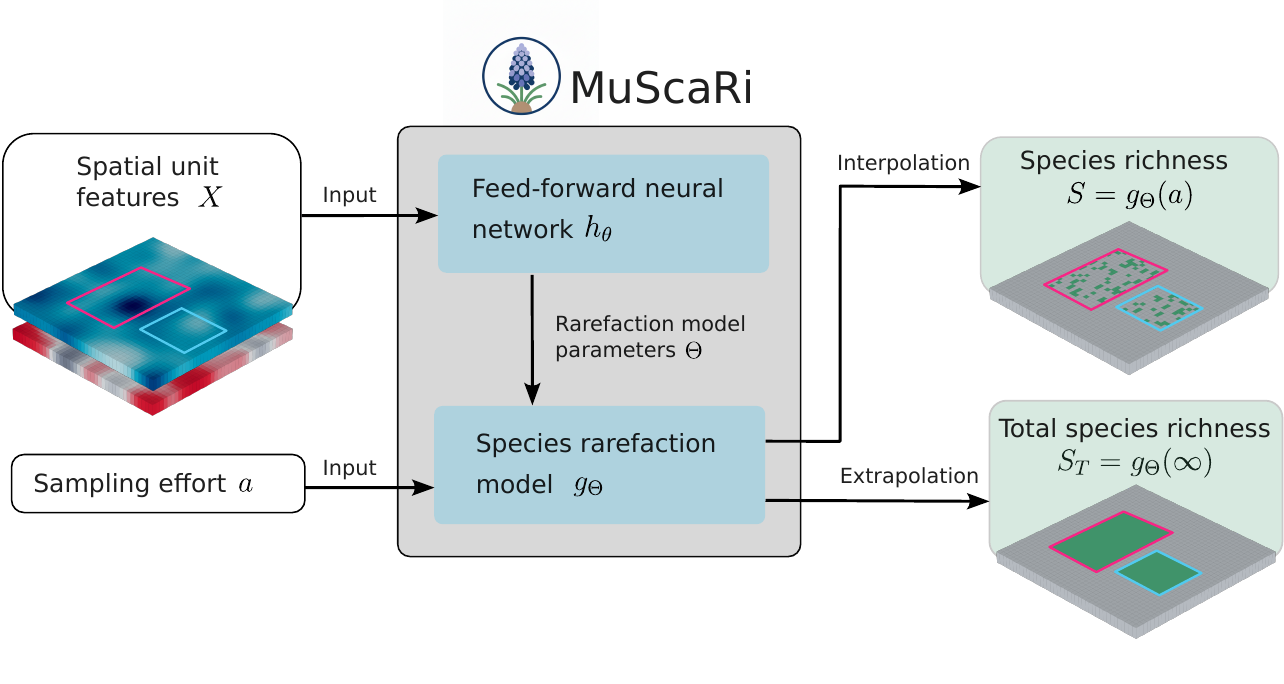}
    \caption{
        \textbf{\modelname{} model architecture.} \modelname{} combines a neural network ($h_\theta$), which maps spatial unit features (denoted as $X$, comprising the spatial unit area and summary statistics of environmental covariates) to the parameters of a theoretically motivated rarefaction model ($g_\Theta$; see \cref{sec:definitions,sec:implementation}). The rarefaction model, in turn, predicts expected species richness as a function of sampling effort $a$ (sum of biodiversity plot areas within the spatial unit).
        \modelname{} is trained end-to-end on samples generated by randomly pooling biodiversity plots within spatial units of varying size and location (blue and red boxes above; see \cref{sec:data}). We say that the model operates in \textit{interpolation mode} when predicting richness at sampling-effort levels observed during training. Because \modelname{} never observes total species richness during training, we say that it operates in \textit{extrapolation mode} when predicting under asymptotic sampling effort ($a\to\infty$), yielding estimates of total species richness ($S_T$).
    }
    \label{fig:architecture}
\end{figure}

Here, we present \modelname{}\footnote{For ``Multi-scale Richness estimation'', also after the \emph{Muscari} genus of perennial bulbous plants.}, a deep learning approach for estimating species richness across spatial scales from small-scale biodiversity plots. For a given location and spatial unit, \modelname{} predicts either the expected species richness at a specified sampling effort or total species richness under asymptotically large sampling effort.
\modelname{} combines a neural network with a parametric rarefaction model: the neural network maps the spatial unit's area and environmental features (hereafter referred to as spatial unit features) to the parameters of the rarefaction model, which in turn predicts expected species richness as a function of sampling effort (see \cref{fig:architecture} and \cref{sec:implementation} for details). The model is trained end-to-end via stochastic gradient descent on subsets of biodiversity plots drawn from randomly placed spatial units of varying area and location (see \cref{sec:data} for details on the training procedure).
Notably, \modelname{} never observes total species richness during training; its capacity to extrapolate beyond sampling-effort levels observed during training is instead conferred by a theoretically motivated rarefaction model, which imposes the necessary monotonicity and saturation constraints (see \cref{sec:definitions} for details).
Richness maps are obtained by predicting total species richness across spatial units of fixed size; querying total species richness for nested spatial units of increasing area at a fixed location yields a SAR curve (see \cref{fig:src_sar}).

We apply \modelname{} to a large dataset from the European Vegetation Archive (EVA, \totkEVAplots{} curated vegetation plots comprising $\sim$9k distinct species; \citealp{chytry2016}) and generate multi-scale estimates of vascular plant species richness across Europe.
To characterize each spatial unit's environmental context, we compute the mean and standard deviation of four bioclimatic variables from the CHELSA dataset \citep{karger2021a,brun2022} and of elevation from the European Digital Elevation Model (EU-DEM; \citep{bashfield2011}) at 1\,km resolution, capturing mean environmental conditions and environmental heterogeneity, respectively.
We benchmark \modelname{} against established baseline richness estimation methods, conduct an ablation study and a Shapley value analysis to disentangle the scale-dependent contributions of area and environmental conditions to richness predictions, and finally generate maps of plant species richness and species accumulation rates across Europe at multiple spatial scales.

\FloatBarrier

\section{Results}

\subsection{Model performance}

The \modelname{} model demonstrated strong predictive performance across evaluation settings, outperforming both a purely data-driven baseline and a classical species richness estimation method. We assessed performance in two complementary modes: interpolation mode, where the model predicts species richness at sampling-effort levels observed during training, and extrapolation mode, where it predicts total species richness under asymptotic sampling effort (see \cref{fig:src_sar} and \cref{fig:architecture}). We compared three \modelname{} variants trained with area only, environmental features only, or both, alongside a feedforward neural network (FFNN) trained with both area and environmental features as a purely data-driven baseline. We also compared \modelname{} models against the Chao2 estimator \citep{chao1987}, a commonly used non-parametric method that estimates total species richness from rare species frequencies. The Chao2 estimator only provides estimates of total species richness (and therefore cannot be used in interpolation mode) and does not leverage environmental context.

\begin{figure}[H]
    \centering
    \includegraphics[width=0.8\textwidth]{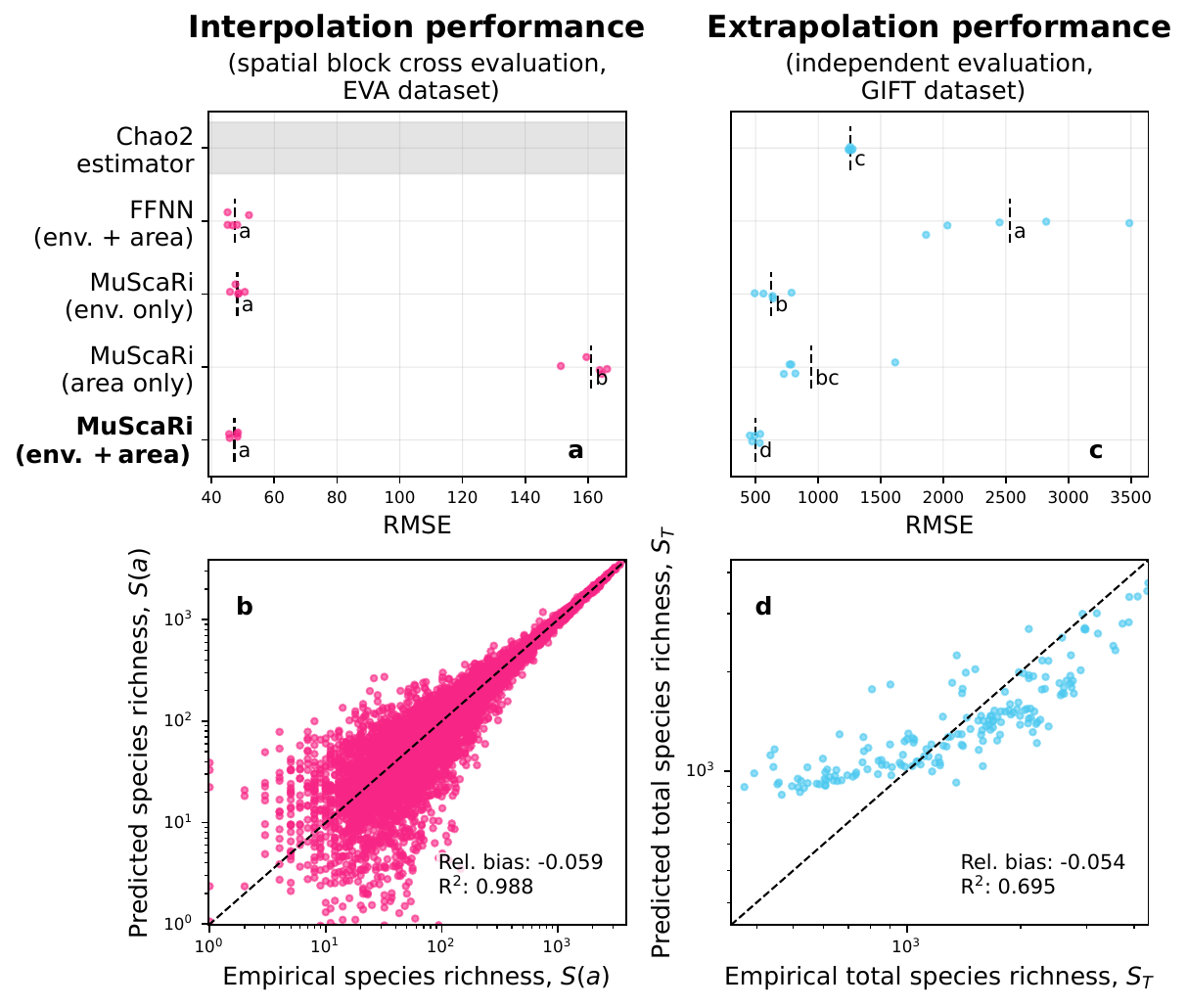}
    \caption{\textbf{Ablation study and model performance}. (\textbf{a}) Interpolation performance (RMSE) for models trained with different feature sets, evaluated on held-out EVA test data at sampling-effort levels observed during training. Each point corresponds to one cross-validation split (see \nameref{sec:methods}); the dashed line indicates the mean across splits. Different lowercase letters indicate significantly different performance (post-hoc Tukey HSD test). (\textbf{b}) Observed versus predicted species richness on a held-out EVA test dataset, from the combined \modelname{} model (area + environment). (\textbf{c}) Extrapolation performance (RMSE on total species richness $S_T$, $a\to\infty$) for models trained with different feature sets, compared against the non-parametric Chao2 estimator and the feedforward neural network baseline (FFNN), evaluated on the GIFT dataset of regional plant inventories. (\textbf{d}) Observed versus predicted total species richness $S_T$ on the GIFT test dataset, from the combined \modelname{} ensemble model (area + environment; see \cref{sec:implementation} for ensembling details).}
    \label{fig:model-perf}
\end{figure}

In interpolation mode, environmental context substantially improved predictive accuracy. Using five-fold spatial block cross-validation (see \cref{sec:data} for details), the \modelname{} variant combining area and environmental features reduced RMSE by 72\% relative to the area-only variant. We found no statistically significant difference in performance between the combined and environment-only variants ($p=0.32$, t-test; \cref{fig:model-perf}\textbf{a}; additional metrics in \cref{tab:interp_performance}). The baseline FFNN model also achieved comparable performance to the combined \modelname{} model ($p=0.85$, t-test), demonstrating that the \modelname{} architecture, and specifically its rarefaction model component (see \cref{sec:implementation}), is sufficiently expressive to capture the complex dependence of richness on sampling effort, area, and environmental conditions.

In extrapolation mode, the combined \modelname{} model substantially outperformed both the baseline FFNN model and the Chao2 estimator. We evaluated predicted total species richness $S_T$ against an independent dataset of exhaustive species inventories: the Global Inventory of Floras and Traits (GIFT; see \cref{sec:data}).
There, the baseline FFNN model performed poorly, with inconsistent performance across folds attributable to undefined behavior outside its training distribution, yielding a mean RMSE approximately $350$\% higher than that of the combined \modelname{} model (\cref{fig:model-perf}\textbf{c}). This result underscores the value of the theoretically motivated rarefaction model when extrapolating beyond observed sampling-effort regimes. The combined \modelname{} model further reduced RMSE by 61\% relative to Chao2 ($p<0.001$, t-test) and yielded substantially less biased richness estimates (median relative bias of $-$5.4\% versus $-$63\% for Chao2), explaining over 68\% of the variance in observed species richness across the GIFT test dataset (see \cref{tab:extrap_performance} and \cref{fig:model-perf}\textbf{d}).
Notably, the combined \modelname{} model outperformed the environment-only variant in this setting, suggesting that conditioning on area is particularly important for extrapolating to large spatial grains, as represented in the GIFT dataset; we investigate this further in the next section.
While the GIFT dataset provides a valuable extrapolation benchmark, it is limited to administrative units that tend to be large (ranging from sub-national regions such as cantons to entire countries, with a median area of 11,695\,km$^2$; see \cref{figSI:datasets}) and whose spatial distribution is geographically biased.
These conclusions held across alternative performance metrics (\cref{tab:extrap_performance}).

Overall, these results confirm that environmental context is essential for predicting species richness across spatial scales, and that \modelname{} effectively leverages small-scale biodiversity plots to deliver improved richness estimates from local to regional scales, outperforming both purely data-driven approaches and non-parametric estimators that do not account for environmental context.
The following sections present a deeper analysis of the combined \modelname{} model (area + environment); unless stated otherwise, \modelname{} hereafter refers to this combined variant.

\subsection{Contribution of area and environment across spatial scales}\label{sec:shapley}

By estimating the contributions of spatial-unit features to \modelname{} model predictions, we can shed light on long-standing macroecological questions. The environmental filtering hypothesis posits that abiotic conditions directly constrain the regional species pool by excluding species whose physiological tolerances are exceeded, such that locations with more favorable conditions support higher species richness \citep{whittaker1972, Vellend2010}. The ``habitat-diversity'' hypothesis holds that environmental heterogeneity promotes higher species richness by increasing niche diversity and supporting greater species accumulation \citep{stein2014, qian2012, turner2005}. The ``passive sampling'' hypothesis predicts that richness increases with area even without added heterogeneity, because larger areas contain more individuals and may bridge dispersal barriers. Finally, the ``area-\textit{per se}'' hypothesis argues that larger areas reduce extinction risk by buffering populations against stochasticity. These mechanisms have been proposed to collectively shape species richness patterns, but their relative importance across spatial scales remains poorly understood \citep{triantis2003, drakare2006}.

To quantify the scale dependence of contributions from environment and area, we computed Shapley values in extrapolation mode for spatial units obtained from a test set from one of the five EVA cross-validation splits (see \cref{sec:data}), grouping environmental features into three predictor classes: mean environmental conditions (mean of environmental covariates), environmental heterogeneity (standard deviation of environmental covariates), and spatial unit area. Shapley values quantify the contribution of each feature to the model output \citep{lundberg2017}. To consistently compare contributions across spatial scales, we computed absolute Shapley values for each predictor class and normalized them to sum to one for each prediction. 

\begin{figure}[H]
    \centering
    \begin{minipage}{0.6\textwidth}
        \includegraphics[width=\textwidth]{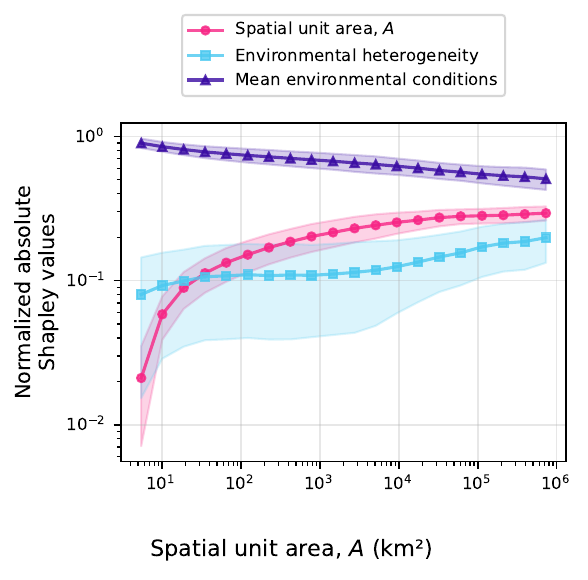}
    \end{minipage}%
    \begin{minipage}{0.3\textwidth}
        \caption{\textbf{Relative importance of area and environmental features in predicting total species richness}. Shapley values indicate the contribution of each predictor class (spatial unit area, mean environmental conditions, and environmental heterogeneity) to total species richness estimates at different spatial grains. Lines show mean normalized absolute Shapley values; shaded areas show standard deviation.
        }
        \label{fig:shapley}
    \end{minipage}
\end{figure}

Results (\cref{fig:shapley}) show greatest support for the environmental filtering hypothesis: mean environmental conditions contributed most to species richness predictions across all spatial scales.
However, as spatial grain increased, the relative contributions of spatial unit area and environmental heterogeneity rose while that of mean environmental conditions decreased, with the increase in area's contribution being more pronounced than that of environmental heterogeneity.
This finding suggests that the passive sampling and area-\textit{per se} hypotheses may be particularly relevant for explaining species richness patterns at large spatial scales.
This scale-dependent shift in area's contribution is also consistent with our ablation study (\cref{fig:model-perf}), where area's effect on model performance became apparent only in the GIFT-based test dataset, which is dominated by large spatial units.
Alternatively, the growing contribution of area at large spatial grains may partly reflect a limitation of the coarse summary statistics used as environmental features: if these statistics fail to fully capture environmental variation at large scales, the model may rely more heavily on area as a proxy for unaccounted environmental heterogeneity.
Although the predictor classes showed only moderate collinearity (see \cref{figSI:corr_mat}), lending confidence to the interpretability of the resulting Shapley values, we recall that Shapley attributions reflect model-based attribution rather than causal decomposition.

Together, these findings highlight the scale dependence of biodiversity drivers \citep{rahbek2005} and quantify the relative contributions of the mechanisms underlying species richness patterns across spatial scales \citep{field2009}.

\subsection{Scale-specific patterns of plant species richness and rate of species accumulation in Europe}\label{sec:maps}

Projections of species richness and species accumulation rates across spatial scales are essential for comparing biodiversity trends from local to regional levels. We used the \modelname{} ensemble model (see \cref{sec:implementation} for details) to map these quantities across Europe at 5 and 50\,km resolution (corresponding to sampling areas of 25\,km$^2$ and 2,500\,km$^2$, respectively), using a moving-window approach. We also estimated location-specific SARs for three environmentally distinct locations (Žďárské vrchy Protected Landscape Area in the Czech Republic, a cool-temperate mixed forest–wetland landscape of moderate relief; Parc Naziunal Svizzer in Switzerland, a high-alpine site with pronounced topographic heterogeneity; and Nationaal Park Veluwezoom in the Netherlands, a low-relief oceanic heathland–forest site on sandy soils) by incrementally increasing the area of spatial units centered at each location.

\begin{figure}[H]
    \centering
    \vspace*{-1.5cm}
    \includegraphics[width=1.0\textwidth, trim=0 2.2cm 0 0, clip]{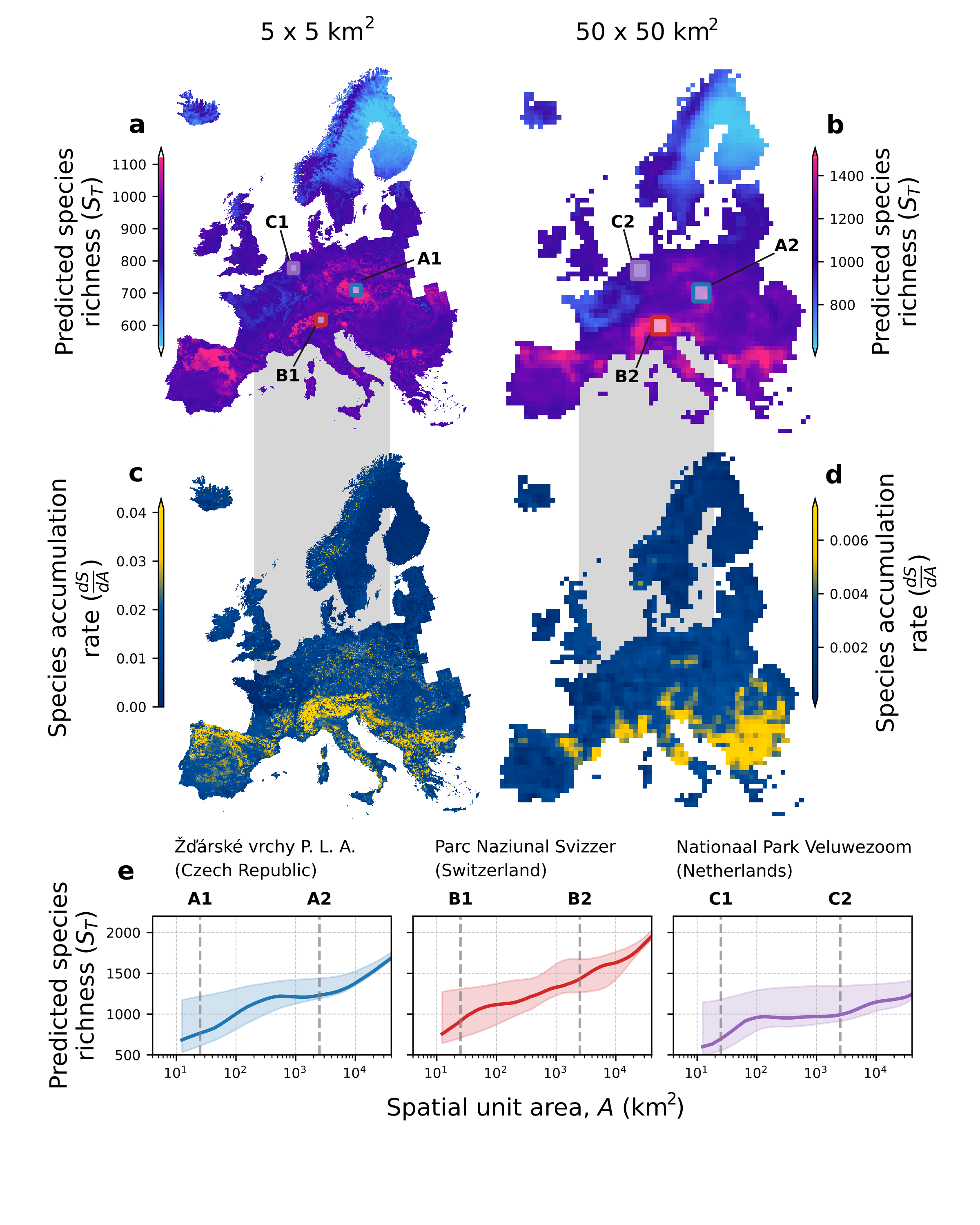}
    \caption{\textbf{Species richness and rate of species accumulation across Europe, predicted at multiple spatial scales.}
    (\textbf{a}-\textbf{b}) Maps of predicted total species richness for square spatial units with edge lengths of 5\,km and 50\,km, respectively, derived as the mean of \modelname{} ensembles (see \cref{sec:implementation} for ensemble details).
    (\textbf{c}-\textbf{d}) Associated maps of species accumulation rates. The species accumulation rate is defined as the local slope of the SAR at the given spatial grain (see \cref{fig:src_sar}).
    (\textbf{e}) Predicted SARs for three environmentally distinct locations: Žďárské vrchy Protected Landscape Area (blue), Parc Naziunal Svizzer (red), and Nationaal Park Veluwezoom (purple), whose positions are indicated in panels (\textbf{a}) and (\textbf{b}). Solid lines represent the ensemble mean; shaded areas indicate one standard deviation of ensemble predictions.
    Additional maps showing the standard deviation of model ensembles are provided in \cref{figSI:maps_std}; numerical values for the three focal locations are given in \cref{tab:sar_summary}.
    }
    \label{fig:maps}
\end{figure}

Predicted species richness maps at 5\,km and 50\,km revealed markedly different richness and accumulation dynamics patterns across geographic regions (\cref{fig:maps}\textbf{a}--\textbf{b}). At 50\,km, patterns were consistent with established biogeographical knowledge of European vascular plant diversity \citep{cai2023}, with the highest richness along the Mediterranean coastline and in mountainous regions including the Alps, Carpathians, and Dinarides. At 5\,km, Central Europe showed comparably high richness values to the Mediterranean and mountainous regions. This may reflect the high local alpha-diversity of Central European mixed forest and grassland communities \citep{vecera2019}, but could also be an artifact of the markedly higher density of vegetation plots in that region in the EVA dataset (see paragraph below, and \cref{figSI:EVA_locations}). Scandinavia showed remarkably similar richness at 5\,km and 50\,km, consistent with the small species pool of boreal flora and the relatively homogeneous environmental conditions across the region \citep{cai2023}. 
These patterns are reflected in the species accumulation rate maps at 5\,km (\cref{fig:maps}\textbf{c}), where the Mediterranean and mountainous regions sustain higher accumulation rates than, e.g., Central Europe. 

The location-specific SARs captured these scale-dependent dynamics (\cref{fig:maps}\textbf{e}). Nationaal Park Veluwezoom showed the slowest accumulation rate, consistent with its homogeneous low-relief sandy heathland–forest landscape (869$\pm$244 species per 25\,km$^2$; 1,102$\pm$166 species per 2,500\,km$^2$; \cref{tab:sar_summary}). Žďárské vrchy showed an intermediate rate, reflecting moderate topographic and habitat diversity across a gently undulating mixed forest–wetland mosaic (901$\pm$232 species per 25\,km$^2$; 1,315$\pm$101 species per 2,500\,km$^2$; \cref{tab:sar_summary}), in close agreement with the 5\,km prediction of an independent study based on a different dataset from Czech flora mapping \citep{klimova2025}. Parc Naziunal Svizzer showed the highest accumulation rate, and consequently the greatest predicted richness at large spatial grains (1,028$\pm$244 species per 25\,km$^2$; 1,508$\pm$162 species per 2,500\,km$^2$; \cref{tab:sar_summary}), attributable to its pronounced topographic heterogeneity and diversity of habitats from subalpine forests to alpine meadows and rocky habitats.

These predictions should be interpreted in light of the standard deviation of the model ensembles (see \cref{figSI:maps_std}), and of the spatial coverage and density of the training data (see \cref{figSI:EVA_locations}). Notably, the standard deviation of the model ensembles decreased with increasing spatial grain (see \cref{tab:sar_summary}), which may reflect the more uniform spatial distribution of vegetation plots within large spatial units, or the fact that species richness becomes more predictable at coarser grains where local stochasticity is buffered and environmental drivers exert a stronger influence \citep{drakare2006}. The model ensembles also yielded large standard deviations in species accumulation rates (\cref{figSI:maps_std,tab:sar_summary}), and may be biased in undersampled regions (e.g., Spain and Scandinavia); predictions in data-sparse regions should therefore be interpreted with particular caution.

\FloatBarrier

\section{Discussion}

We demonstrated how combining deep learning with a theoretically motivated rarefaction model enables spatially explicit prediction of species richness from fine to coarse spatial grains. By conditioning species rarefaction curves on spatial unit area and environmental features, \modelname{} leverages well-sampled regions to generate predictions in environmentally similar but data-sparse areas. Composing a neural network with a parametric rarefaction model that enforces monotonicity and saturation constraints allows \modelname{} to upscale species richness to spatial scales orders of magnitude beyond those of typical biodiversity plots. This contrasts with purely data-driven approaches such as \cite{andermann2022}, which interpolate richness at observed sampling efforts without modeling the underlying accumulation process; like the FFNN baseline in \cref{fig:model-perf}, such approaches perform poorly at estimating total species richness, as doing so requires extrapolating beyond the sampling-effort range seen during training. Against an independent test set of exhaustive regional inventories, \modelname{} explained 68\% of the variance in total species richness, reducing RMSE by 61\% relative to the Chao2 estimator \citep{chao1987} and yielding substantially less biased predictions. A further advantage is its capacity to deliver richness estimates at arbitrary spatial scales within a single model: in contrast to \citet{andermann2022}, who trained three separate models for $\alpha$-diversity, $\gamma$-diversity, and species turnover at fixed scales, \modelname{} accommodates all three metrics as special cases of the same predictive framework, making it a versatile tool for applications that require consistent richness estimates across spatial units of varying size, such as cross-regional biodiversity assessments.

Machine learning applications in macroecology have predominantly focused on predicting species distributions (see, e.g., \cite{Deneu2021,brun2024,gillespie2024}), yet many ecological and conservation questions require community- or ecosystem-level properties. \modelname{} takes a complementary approach by directly predicting species richness rather than explicitly modeling individual species distributions, which confers two advantages over joint species distribution models (J-SDMs). First, \modelname{} is more data-efficient: while J-SDMs can borrow statistical strength across species to improve predictions for rare species, they must still estimate species-specific parameters for each species in the community, with estimation errors potentially accumulating with species number and propagating into biased estimates of total richness. By predicting richness directly, \modelname{} bypasses these issues and can readily accommodate species-rich communities dominated by rare species. Second, by learning a macroscopic representation of ecosystem state, \modelname{} can capture emergent community-level properties arising from assembly processes that J-SDMs cannot resolve \citep{leroux2017, dubuis2011}. An interesting avenue for future work would be to combine J-SDMs with \modelname{} to predict realistic species assemblages \citep{dubuis2011}.

The theoretical shape of the SAR has long been debated in ecology, with no consensus on a single best model \citep{dengler2009}. While the search for a theoretical functional form is required in a low-data regime or for deriving general macroecological patterns, theoretical SAR models are simplifications that cannot accommodate the complexity of realistic landscapes. Realistic landscapes are often heterogeneous, where species richness may rise idiosyncratically due to dispersal barriers, natural local variations, or anthropogenic stressors \citep{whittaker1972, gering2003}.
Our approach allows leveraging abundant biodiversity plot data, which, combined with the proposed \modelname{} architecture and training strategy, has the capacity to capture these complex accumulation dynamics without imposing strong assumptions about the expected SAR functional form.

Beyond richness estimation, \modelname{} provides an essential tool for biodiversity assessment and conservation planning. For instance, it could be used to assess the impact of habitat loss by comparing predicted species richness between intact and degraded areas \citep{gotelli2001, matthews2021, thomas2004, brooks2002}.
Species accumulation rate maps are particularly valuable in this context: high-accumulation rate regions indicate that habitat loss may disproportionately impact regional biodiversity by eliminating compositionally distinct assemblages \citep{socolar2016}, whereas areas with high local richness but low accumulation rates may show greater resilience to localised habitat loss. 
Overall, \modelname{} delivers information on multi-scale biodiversity trends that may be used to address core needs in biodiversity conservation \citep{kunin2018}.

We conclude by identifying the most important limitations and avenues for future work.
First, the EVA dataset is characterized by low vegetation plot density in certain regions (see \cref{figSI:EVA_locations}), which may introduce prediction bias or high uncertainty in those regions. The ensemble model enabled partial quantification of this uncertainty (see \cref{figSI:maps_std}), but more comprehensive bias and uncertainty characterization represents an important avenue for ensuring transparent biodiversity assessments.
Second, our approach relies on the assumption that biodiversity plots are uniformly placed within spatial units, but in reality, surveys often concentrate along accessible areas such as roads or trails. Furthermore, we assume that equal-area sampling yields comparable information across diverse plot combinations; yet, units with a few large plots may not be directly comparable to those with many small plots. Non-uniform plot placement and differences among equal-area plot combinations are expected to bias species richness estimates. This could be addressed in future work by explicitly modeling the spatial distribution of the sampling effort within spatial units.
Third, our framework does not consider the temporal dimension of biodiversity records, whereas evaluating the state and trends of biodiversity under global change ultimately requires temporally-resolved estimates \citep{ipbes_global_2019}. Future work could relax this stationarity assumption on the underlying species pool dynamics and treat time more explicitly to detect directional trends in species richness driven by land-use change or climate shifts.
Fourth, the simplistic summary statistics (mean and standard deviation) used to characterize environmental conditions within spatial units may not fully capture the underlying environmental context. Future work could address this by using convolutional neural networks to automatically extract higher-order statistics of environmental variables and characterize more subtle drivers of species richness, such as habitat fragmentation effects \citep{hanski2013, rybicki2013}. These models could also process satellite imagery, from which we anticipate that fine spatial details could further improve model performance, particularly at fine spatial grains.
Fifth, the GIFT dataset used to evaluate extrapolation performance comprised a small number of spatial units whose areas were biased toward large extents (median area of 11,695\,km$^2$; see \cref{figSI:datasets}). The scarcity of exhaustive species inventories limits our understanding of model behavior in extrapolation mode at finer spatial grains.
Sixth, our model predicted vascular plant species richness without distinguishing between habitat types. Future work could specialise the model to predict habitat-specific richness by training on relevant plant community subsets, or extend it to predict species richness for other taxonomic groups.

\section{Conclusion}

Our study demonstrates how deep learning combined with sampling theory can effectively estimate macroecological properties of ecosystems from opportunistic biodiversity plot datasets. \modelname{} provides a data-efficient approach to deliver comprehensive insights into biodiversity patterns, revealing the distinct effects of biodiversity drivers across scales. We anticipate future applications of our framework to capture complementary macroecological properties of ecosystems, such as functional or genetic diversity. The multi-scale prediction capacity of \modelname{} is essential for attributing biodiversity changes to global change drivers and designing robust conservation strategies that account for the scale dependency of ecological processes.

\section{Methods}\label{sec:methods}
\newcommand{\ma}{\mathsf{a}} 
\newcommand{\mA}{{\mathcal{A}}} 
\newcommand{\is}{{\boldsymbol{i}}} 
\newcommand{\cS}{\mathcal{S}} 

We aim to estimate \emph{total species richness} $S_\mathrm{T}$, the total number of species in a spatial unit $\mA$ at a given location (vascular plant species richness, in the specific case of this manuscript).
However, biodiversity surveys typically sample only a small fraction of $\mA$, requiring extrapolation from these limited samples to estimate total species richness under asymptotic sampling effort. This extrapolation challenge has been addressed through various approaches \citep{elenaschmitz2025}, but these methods typically focus on estimating total richness within a single spatial unit.
Our approach introduces a key innovation by conditioning a parametric rarefaction model (characterising the rarefaction curve of a spatial unit) on environmental features of that unit, including its area and summary statistics of environmental covariates. This conditioning enables spatially explicit mapping of $S_\mathrm{T}=S_\mathrm{T}(X)$ across a geographic region at multiple spatial scales.

\Cref{sec:definitions} provides definitions, explains our methodological setup, and theoretically motivates the chosen rarefaction model functional form. \Cref{sec:implementation} details the curve fitting procedure and neural network implementation. \Cref{sec:data} describes the datasets used in our analysis, and the training, validation and test sample generation procedure.

\subsection{Estimating total species richness from small samples}\label{sec:definitions}

\paragraph{Sampling process and sampling effort.} We define a biodiversity plot set as a collection of $m=|\is|$ small-scale biodiversity plots
$\ma_{i_1},\ldots,\ma_{i_m}$ placed within the spatial unit $\mA$ with area $A=|\mA|$:
\begin{equation}
  \ma_\is := \{\ma_{i_1},\ldots,\ma_{i_m}\}, \qquad \bigcup_{i\in\is} \ma_i \subseteq \mA,
\end{equation}
where the biodiversity plots have corresponding locations $x_{i_1}, ..., x_{i_m}$ and areas $|\ma_{i_1}|,\ldots,|\ma_{i_m}|$. We will view a particular $\ma_\is$ 
as a realization of a random variable. We quantify \emph{sampling effort} within $\mA$ via the sum of the sample areas:
\begin{equation}
    a_{\is} := \sum_{i\in\is} |\ma_i|,\label{eq:sumeffort}
\end{equation}
where $a_{\is}$ is unbounded due to overlaps at high effort. 
In (the union of) $\ma_\is$, we count species to obtain its \emph{observed species richness}:
\begin{equation}\label{eq:Sobs}
	S(\ma_\is):= \sum_{j=1}^{S_\mathrm{T}} \mathds{1}_{ N_j(\ma_\is)\geq 1},
\end{equation}
where $\mathds{1}_{(\cdot)}$ evaluates to 1 when the condition in its argument is true and to zero otherwise, and $N_j(\ma_\is)$ the
number of individuals of species $j$ in $\ma_\is$. Repeatedly sampling over different subests of biodiversity plots yields effort-richness pairs $a_{\is} \mapsto S(\ma_\is)$. 

\paragraph{Species rarefaction curve.} 
The curve that shows how \emph{expected species richness} increases with sampling effort is known as a rarefaction curve. We can obtain it by taking an expectation of species richness conditional on sampling effort
\begin{equation}
	\cS(a) := \E\left[S(\ma_\is) \,\Big|\, a_\is=a\right]
	=\sum_{j=1}^{S_\mathrm{T}}\mathbb{P}\left[N_j(\ma_\is)\ge1 \,\Big|\, a_\is=a\right],
\end{equation}
where we used \cref{eq:Sobs} and that the expectation of an indicator function is the probability of its argument.
Defining
\begin{equation}\label{eq:sad}
	p_n(a):=\frac{1}{S_\mathrm{T}}\sum_{j=1}^{S_{\mathrm T}} \mathbb{P}\left[N_j(\ma_\is)=n \,\Big|\, a_\is=a\right],
\end{equation}
the expected proportion of species in a sample with $n$ individuals
at effort $a$, we obtain
\begin{equation}
	\cS(a) = S_\mathrm{T}\sum_{n=1}^\infty p_n(a)=S_\mathrm{T}[1-p_0(a)],
\end{equation}
where we used that \cref{eq:sad} sums to 1. As $S_T$ denotes the number of species occurring in $\mA$, $p_0(a)$ is the expected fraction of species occurring in $\mA$ that are missed at effort $a$. 
We can write $p_0$ and therefore $\cS(a)$ as (see SI \Cref{sec:decp0}; \citet{harte2009})
\begin{equation}
	p_0(a)=\sum_{n\ge 1}f^+_n q_n(a),\quad\Rightarrow\quad 	\cS(a) = S_\mathrm{T}[1-\sum_{n\ge 1}f^+_n q_n(a)].\label{eq:p0}
\end{equation}
where $f^+_n$ is the zero-truncated species abundance distribution (SAD, over species present) in $\mA$ and $q_n(a)$ the probability of 
absence at effort $a$ given $n$ individuals in $\mA$. $q_n$ depends on how individuals are arranged and sampled within $\mA$.
By construction, $\mathcal{S}(0)=0$, $\mathcal{S}\to S_\mathrm{T}$ for $a\to\infty$, and $\mathcal{S}$ is monotonically increasing in $a$.

\paragraph{Parametric estimation of the rarefaction curve.}
The rarefaction curve thus depends jointly on the SAD, the spatial arrangement of individuals within $\mA$, and the sampling process. Previous work showed that closed form richness-area curves can be obtained when assuming
uniform random placement (URP) of individuals \citep{Coleman1981,he2002}. Under URP, independence permits writing $q_n=q_1^n$, such that \cref{eq:p0} becomes
\begin{equation}\label{eq:cS}
	\cS_\vartheta(a) =S_{\mathrm{T}}\bigl[1-G\{f^+_{n,\vartheta}\}\bigl(q_1(a)\bigr)\bigr],
\end{equation}
where $G\{f^+_{n,\vartheta}\}(q_1):=\sum_{n\ge1}f^+_{n,\vartheta}q_1^n$ is the probability-generating function (PGF) of the SAD with parameters $\vartheta$, and $q_1$ depends on the sampling process  (see \cref{sec:URP}, for more detail). For instance,
when $f^+_{n,\vartheta}$ is a geometric distribution with parameter $\vartheta$,
then the rarefaction curve arising from our sampling process is $\cS_\vartheta(a)=S_\mathrm{T}\tfrac{1-\exp(-a/|\mA|)}{1-(1-\vartheta)\exp(-a/|\mA|)}$.
This suggests that we can estimate $\cS_\vartheta(\cdot)$ from effort-richness pairs by first choosing the right parametric family of probability mass functions $f^+_{n,\vartheta}$ with parameter(s) $\vartheta$, then obtaining the corresponding form of the rarefaction curve
via \cref{eq:cS}, and finally obtaining the parameters $\Theta=(S_\mathrm{T},\vartheta)$ by fitting it to the data. 

Real communities, however, exhibit spatial structure due to environmental heterogeneity and spatial interactions, so URP is typically violated. Moreover both spatial structure and the SAD generally differ between 
locations, so it would be highly restrictive to choose a single parametric family of probability mass functions. Hence, any chosen functional form $g_\Theta(a)$ must be flexible enough as to fit diverse shapes regardless of the deviation from URP and differences between locations, but still satisfy monotonicity in $a$ and reach $S_\mathrm{T}$ for large sampling effort. 
Parameters $\Theta$ can be obtained by fitting $g_\Theta$ to effort-richness data pairs $a_\is \mapsto S(\ma_\is)$, generated through the sampling process. Fitting can be done via maximum likelihood estimation. Once $g_\Theta$ is fitted, we can recover $S_T$ as its large sampling effort limit.

\paragraph{Four-parameter Weibull function.} A flexible choice for the functional form $g_\Theta$ that meets the required constraints is the four-parameter Weibull function:
\begin{equation}\label{eq:weibullfit}
	g_\Theta(a) = c + (S_\mathrm{T}-c) (1-\exp\left[- (\ln a/e)^b \right]),
\end{equation}
This function is positive, monotone and saturating provided $S_T>c$, $b,e>0$ and $a\ge1$. It is at its minimum ($c$) at $a=1$ and converges to its maximum ($S_\mathrm{T}$) as $a\to\infty$.
The offset $c$ captures the species richness at $a=1$. 
$b$ is the shape parameter controlling the rate of species accumulation, and $e$ is the scale parameter representing the characteristic sampling effort. This species rarefaction model, previously used in, e.g., \cite{zou2023}, allows for a wide range of rarefaction curve shapes, while ensuring the necessary saturation and monotonicity constraints.

\subsection{\modelname{} model}\label{sec:implementation}
\paragraph{Conditioning the rarefaction curve on spatial unit features.}
Rarefaction curves depend in general on the area and the environmental covariates associated with the spatial unit under consideration \citep{dengler2010}. 
To predict the observed species richness $S$ of a spatial unit $\mA$ with arbitrary size and location, we condition $g_\Theta$ on features $X$ associated with the spatial unit. We use a neural network $h_\theta$ to learn the mapping from the features $X$, which may include the spatial unit area $|\mA_k|$, and/or other summary statistics of environmental covariates, to the parameters $\Theta$ of the rarefaction curve model $g_\Theta$:
\begin{equation}
    \Theta = h_\theta(X)
\end{equation}
We refer to $a, X \mapsto \M_\theta = g_{h_\theta(X)}(a)$ as the \modelname{} model, which takes as input both the sampling effort $a$ and features $X$. 
\modelname{} is trained end-to-end by minimizing the loss function:
\begin{equation}\label{eq:loss}
        L(\theta) = \sum_{k=1}^K \left[\M_\theta(a_{\is_k}, X_k) - S(\ma_{\is_k})\right]^2,\quad \bigcup_{j\in \is_k} \ma_j\subset\mA_k
\end{equation} 
via stochastic gradient descent, where $S(\ma_{\is_k})$ represents the observed species richness at effort $a_{\is_k}=\sum_{j\in\is_k} |\ma_j|$ from the randomly selected biodiversity plots $\ma_{\is_k}$ located within spatial unit $\mA_k$ with the associated features $X_k$.
We detail the generation of the triplets ($a_{\is_k}$, $X_k$, $S(\ma_{\is_k})$) in \cref{sec:data}.

\paragraph{Predicting rarefaction curves, total species richness and SARs.}
Once trained, the \modelname{} model can be thought of as an approximation of the expected species richness conditioned on sampling effort and environmental features:
\begin{equation}\label{eq:deepSAR}
    \mathbb{E}[S(a)\mid X] \approx \M_\theta(a, X)
\end{equation}
The \modelname{} model can therefore predict rarefaction curves for any spatial unit $\mA$ with associated features $X$ by varying $a$ while keeping $X$ fixed.
Importantly, the \modelname{} model can also predict total species richness ${S_\mathrm{T}}$ for any spatial unit $\mA$ with associated features $X$ by evaluating $\M_\theta$ at infinite sampling effort:
\begin{equation}
    {S_\mathrm{T}} = \lim_{a\to\infty} \M_\theta(a, X).
\end{equation}
In practice, we obtain ${S_\mathrm{T}}$ by directly retrieving ${S_\mathrm{T}}$ from the neural network output $h_\theta(X)$.
SARs can be obtained by computing $S_\mathrm{T}$ for spatial units centered at the same location but with increasing areas (nested spatial units, see \cref{fig:src_sar}).

\paragraph{\modelname{} ensemble model.}
For ensemble predictions, we used the five models trained on the five independent cross-validation splits (see \cref{sec:data}), whose predictions were aggregated using the arithmetic mean. Standard deviations of ensemble predictions in \cref{fig:maps} are provided in \cref{figSI:maps_std}, and can serve as an estimate of uncertainty (model-based uncertainty; \cite{gal2016}).

\paragraph{Neural network architecture and hyperparameters.}
Both \modelname{} and the purely data-driven baseline employ feedforward neural networks (FFNNs) implemented in PyTorch. Both neural networks share the same base architecture of 6 fully connected layers with 32, 128, 256, 256, 128, and 32 neurons and leaky ReLU activations, except for the last layer. Both neural networks take as input spatial unit features $X$.
\modelname{} and the purely data-driven models differ in how sampling effort $a$ enters the prediction and in their output layers. In \modelname{} models, $a$ is injected directly into $g_\Theta$ rather than passed to $h_\theta$. The last layer of $h_\theta$ outputs the four parameters of \cref{eq:weibullfit}, enforcing $c > d$ and $e \geq 0$ (see source code for details). In the purely data-driven model, $a$ is concatenated with $X$ as an additional input. The last layer applies an exponential activation to output richness directly, and batch normalisation is applied after each hidden layer, which we found to improve convergence.

\subsection{Data}\label{sec:data}
\paragraph{Species richness data.}
We used a dataset of 502,724 vegetation plots from the European Vegetation Archive (EVA, \cite{chytry2016}; \url{https://doi.org/10.58060/d1bp-fp47}; project 172; data retrieved on 27 February 2023). We selected only plots with available plot area information and coordinate uncertainty less than 1 km. We additionally discarded plots not located on land and plots recorded prior to 1972, and excluded non-vascular taxa from the dataset. Vegetation plots were analyzed without distinction of habitat types. This data filtering step resulted in \totEVAplots{} vegetation plots for downstream analyses (see \cref{figSI:EVA_locations}).
Additionally, we used the Global Inventory of Floras and Traits (GIFT) dataset \citep{weigelt2020}, filtered to select inventories occurring within the geographic range of the EVA vegetation plots, resulting in 184 exhaustive species inventories.
We harmonized species names between the EVA and GIFT datasets following procedures detailed at \url{https://github.com/vboussange/MuScaRi}, and restricted species in the GIFT dataset to those in the EVA dataset. This resulted in both EVA-based and GIFT-based datasets comprising a total of \totSpecies{} distinct species. For training, we normalized species richness values by the maximum observed richness in the training set.

\paragraph{Environmental features.}
We calculated environmental features from four bioclimatic variables of the CHELSA dataset \citep{karger2021a,brun2022}: mean annual air temperature (\texttt{bio1}), annual precipitation (\texttt{bio12}), near-surface wind speed (\texttt{sfcWind}), and potential evapotranspiration (\texttt{pet}); and from elevation data obtained from the European Digital Elevation Model (EU-DEM; \citep{bashfield2011}). All rasters were resampled to a common resolution of 1\,km. For each spatial unit, we computed the mean and standard deviation of each covariate over the raster cells it intersects, yielding features that characterize mean environmental conditions and environmental heterogeneity, respectively. Pairwise correlations among features belonging to different predictor classes (spatial-unit area, mean of covariates, and standard deviation of covariates) were moderate (absolute Spearman $|\rho| < 0.6$; see \cref{figSI:corr_mat}), supporting meaningful ablation and Shapley value analyses. All features were normalized to the range $[0, 1]$ using statistics computed from the training set.

\paragraph{Sample generation and spatial block cross-validation procedure with the EVA dataset.}
We generated sample sets $\ma_{\is_k}$ (see \cref{eq:loss}) by randomly placing spatial units $\mA_k$ that contain at least one vegetation plot. Spatial-unit areas $|\mA_k|$ were drawn from a log-uniform distribution over $[4, 10^{6}]$\,km$^2$; the log-uniform prior ensures balanced coverage across spatial scales, compensating for the decreasing variance in species richness at large areas as richness converges toward the regional species pool. The lower bound ensures that no spatial unit falls below the resolution of the environmental covariates, and the upper bound ensures that training areas span and exceed those represented in the GIFT test dataset. Spatial units were allowed to overlap.
Within each spatial unit $\mA_k$, we identified the $M_k$ EVA plots whose coordinates fall within its bounds. We then selected a random subset $\ma_{\is_k}$ of these plots, with the number of plots $|\is_k|$ drawn from a log-uniform discrete distribution $|\is_k| \sim \mathcal{U}_{\log}\{2, 3, \ldots, M_k\}$, so that all sampling-effort scales are represented equally, and computed the observed species richness $S(\ma_{\is_k})$ by counting the total number of unique species across the selected plots. The resulting triplets $(a_{\is_k}, X_k, S(\ma_{\is_k}))$ serve as inputs and labels for training and evaluating the models.

For fair model evaluation, we generated five train, validation, and test splits using a spatial-blocking procedure applied at the level of individual vegetation plots. We divided Europe into spatial blocks of 1\,km\,$\times$\,1\,km and assigned each block to one of five folds; each plot inherits the fold of the block containing it. This block size was chosen to ensure approximately uniform plot density across folds. For each split, one fold served as the test-plot set, one as the validation-plot set, and the remaining three as the training-plot set. Spatial units were then generated independently for each train, validation, and test set using only the plots from their respective fold. This design ensures that each plot appears in exactly one test set and prevents data leakage within a split even when spatial units from different sets overlap geographically. We show the spatial distribution of the EVA vegetation plots and the train, validation, and test sets of one split in \cref{figSI:spatial_blocks}.
We assessed how the number of training samples (triplets ($a_{\is_k}$, $X_k$, $S(\ma_{\is_k})$)) influences model performance and report the results in \cref{figSI:scaling_exp}. We found that model performance stabilised when the number of training samples was on the order of magnitude of the total number of plots.
The exact number of training, validation, and test samples used within each split is reported in \cref{tab:fold_sample_counts}.

\paragraph{Test sample set generation with the GIFT dataset.}
To evaluate the model's capacity to estimate total species richness by extrapolation under asymptotic sampling effort, we used the GIFT dataset \citep{weigelt2020}. For each entry, corresponding to an exhaustive species inventory of a political unit (mostly countries, but also regions), we derived environmental features $X_k$; the recorded total species count ${S_T}_k$ served as the ground truth label for the model's prediction $\lim_{a\to\infty} \M_\theta(a, X_k)$.

\section{Data availability}
The curated EVA dataset with anonymized species names, the associated curated GIFT dataset anonymized following the same procedure, the preprocessed environmental covariates, and usage tutorials are available at \url{https://huggingface.co/datasets/vboussange/muscari-data}. The pretrained weights of the \modelname{} ensemble model and usage tutorials are available at \url{https://huggingface.co/vboussange/muscari}. The original EVA dataset, archived at \url{https://doi.org/10.58060/d1bp-fp47}, has restricted access and is available upon request following the guidelines at \url{https://euroveg.org/eva-database}. Scripts for downloading and processing the original GIFT dataset are provided in our GitHub repository. Original environmental covariates are available from the CHELSA database at \url{https://chelsa-climate.org}. Original elevation data from EU-DEM are available at \url{https://ec.europa.eu/eurostat/web/gisco/geodata/digital-elevation-model/eu-dem}.

\section{Code availability}\label{sec:code_availability}
The code used to preprocess the original data, train and evaluate the models and generate all figures is available under our GitHub repository at \url{https://github.com/vboussange/MuScaRi}.

\section{Author contributions}
V.B., B.W., P.B., J.T.M., G.M., J.P., T.S., N.E.Z., and D.N.K. conceived the study. V.B. developed the methodology and implemented the software. V.B. and B.W. performed the formal analysis. P.B., J.T.M., G.M., J.P., T.S., and D.N.K. contributed to methodological development. Data were provided by I.A., H.B., M.C., S.K., Z.L., M.V., I.B., K.T.E., J.L., and J.-C.S. V.B. wrote the original draft. All authors contributed to manuscript review and editing. N.E.Z. and D.N.K. acquired funding. D.N.K. supervised the project.
Authors are listed alphabetically in the following order: the core team leading the study (V.B., P.B., J.T.M., G.M., J.P., T.S., N.E.Z.), followed by contributors from the FeedBaCks consortium (I.A., H.B., M.C., S.K., Z.L., M.V.) and EVA data custodians (I.B., K.T.E., J.L., J.-C.S.), also alphabetically within each group, and last the senior author (D.N.K.).

\section{Acknowledgements}
This research was funded through the 2019-2020 BiodivERsA joint call for research proposals, under the BiodivClim ERA-Net COFUND program, and with the funding organisations Swiss National Science Foundation SNF (project: FeedBaCks, 193907), the German Research Foundation (DFG BR 1698/21-1, DFG HI 1538/16-1), and the Technology Agency of the Czech Republic (SS70010002). N.E.Z. considers this work a contribution to the SPEED2ZERO Joint Initiative, which received support from the ETH-Board under the Joint Initiatives scheme.
J.-C.S. considers this work a contribution to Center for Ecological Dynamics in a Novel Biosphere (ECONOVO), funded by Danish National Research Foundation (grant DNRF173).

\printbibliography

\appendix
    \renewcommand{\thetable}{S\arabic{table}}
    \renewcommand{\theequation}{S\arabic{equation}}
    \renewcommand{\thefigure}{S\arabic{figure}}
    \renewcommand{\thesection}{S\arabic{section}}
	\setcounter{equation}{0}
    \setcounter{figure}{0}
    \setcounter{table}{0}
    \include{SI}
    \section{Supplementary text}
    \subsection{Decomposition of $p_0$}\label{sec:decp0}
    
    \paragraph{Step 1. Definition.}
    \begin{equation}
    	p_0(a):=\mathbb{P}\!\left(N(a)=0 \,\mid\, N(\mA)\ge 1\right).
    	\label{eq:defP0cond}
    \end{equation}
    
    \paragraph{Step 2. Law of total probability for conditional probabilities.}
    \begin{equation}
    	\mathbb{P}\!\left(N(a)=0 \,\mid\, N(\mA)\ge 1\right)
    	=\sum_{n\ge 1}\mathbb{P}\!\left(N(a)=0 \mid N(\mA)=n\right)\;
    	\mathbb{P}\!\left(N(\mA)=n \mid N(\mA)\ge 1\right).
    	\label{eq:lota_cond}
    \end{equation}
    
    \paragraph{Step 3. Identify ZT–SAD and spatial factor.}
    \begin{equation}
    	f_n^{+}\;:=\;\mathbb{P}\!\left(N(\mA)=n \mid N(\mA)\ge 1\right),\qquad
    	q_n(a)\;:=\;\mathbb{P}\!\left(N(a)=0 \mid N(\mA)=n\right),
    \end{equation}
    so
    \begin{equation}
    		p_0(a)\;=\;\sum_{n\ge 1} f_n^{+}q_n(a).
    	\label{eq:decomp_cond}
    \end{equation}

    \section{Closed forms under uniform random placement} \label{sec:URP}

\begin{figure}[h!] 
        \centering
        \includegraphics[width=1.\textwidth]{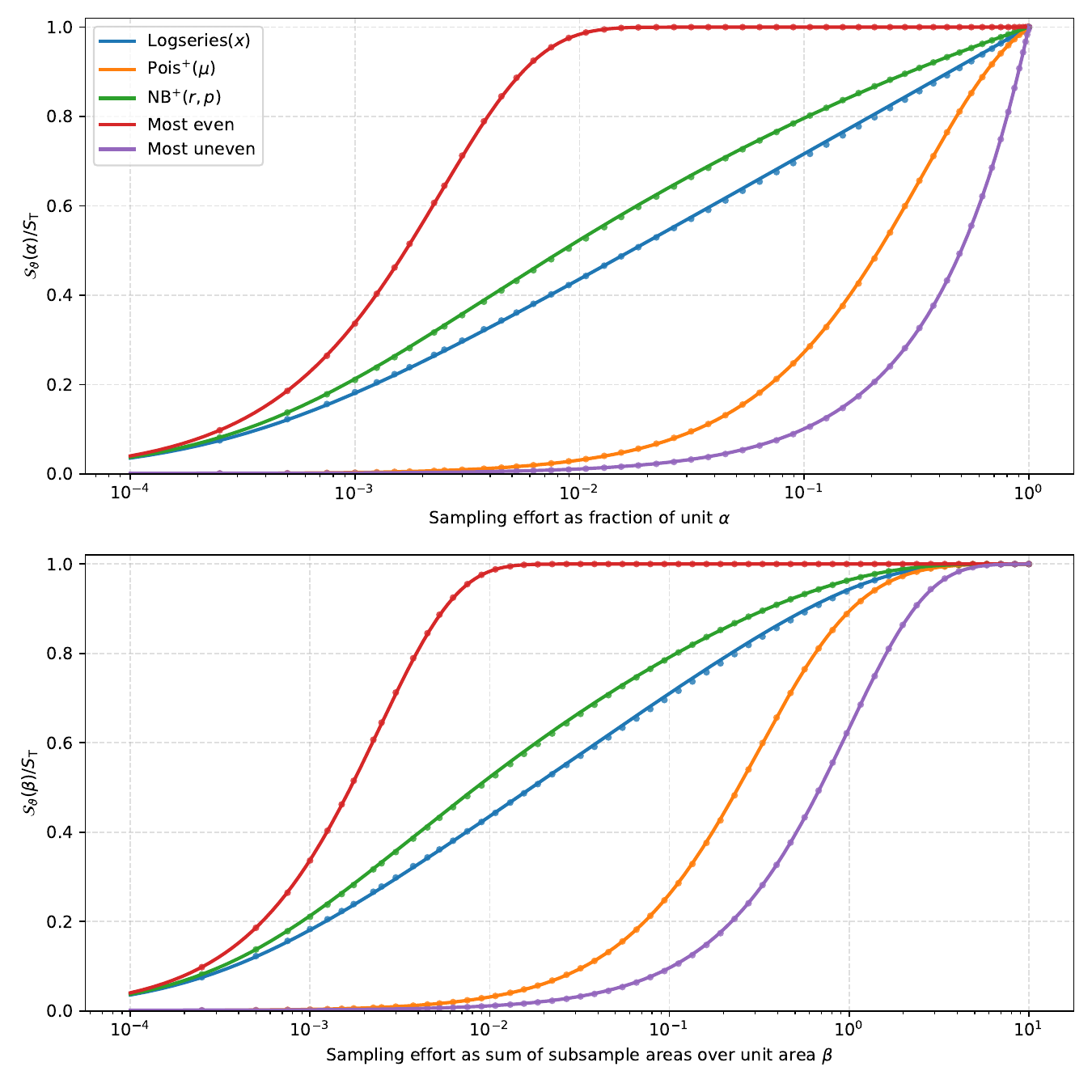}
        \caption{\textbf{Uniform random placement (URP) rarefaction: Monte Carlo check vs theory.}
        Individuals of $S_{\mathrm{T}}=814$ species are placed independently and uniformly at random on a $400\times 250$ periodic grid with their frequencies determined by the SADs of \cref{tab:ESA_binom_SADs}. Sampling consists of adding random $5\times 5$ plots; effort is measured either by the realized union area fraction (top) or by the summed area fraction (bottom). Solid curves show the theoretical expectation $\mathcal{S}_\vartheta(\cdot)/S_{\mathrm{T}}$ for the five SADs shown in \cref{tab:ESA_binom_SADs}. Points show Monte Carlo means over many realizations.}
         \label{figSI:rarefactionURP}
    \end{figure}

    Here, we show that under uniform random placement of individuals (URP), closed form
    solutions for the rarefaction curve can be obtained. This was shown earlier by e.g.
    \citet{Coleman1981,he2002}, yet using the union of sampled areas as
    effort quantity. As we permit sampled areas to overlap, our quantity of effort is
    the sum of sampled areas. 
    The resulting closed forms differ, but both cases can be written in the common form:
    \begin{equation}
	   \cS_\vartheta(a) =S_{\mathrm{T}}\bigl[1-G\{f^+_{n,\vartheta}\}\bigl(q_1(a)\bigr)\bigr].
    \end{equation}
    where $G\{f^+_{n,\vartheta}\}(q_1):=\sum_{n\ge1}f^+_{n,\vartheta}q_1^n$ is the probability-generating function (PGF) of the SAD with parameters $\vartheta$, $q_1(a)$ is the probability that a single individual is not captured at effort $a$ under the chosen sampling process,  and
    $a$ is the effort quantity consistent with the sampling process. The step that permits this form is the substitution of $q_n=q_1^n$ in
    \cref{eq:p0}, which follows from independence.  
    We explain both cases below. Examples for specific choices of the SAD are shown
    in \cref{tab:ESA_binom_SADs} and \cref{figSI:rarefactionURP}.
    
    \paragraph{Richness versus union of sampled areas.}
    For a biodiversity plot set $\ma_\is=\{\ma_i\}_{i\in\is}$, define union-based effort as the union area
    \begin{equation}
        a_{\is} := \left|\bigcup_{i\in\is} \ma_i\right|. \qquad
    \end{equation}
    Under URP, the probability that any specific individual falls within the sampled
    region does not depend on $\is$ and is simply the fraction of area covered, $a/|\mA|\in [0,1]$. Consequently,
    the number of individuals $N(a)$ sampled in $a$ given a total abundance $n$ in $\mA$ follows
    a Binomial distribution:
    \begin{equation}
        N(a)\mid N(\mA)=n \sim \mathrm{Bin}\!\left(n,\frac{a}{|\mA|}\right).
    \end{equation}
    The probability of missing a species with $n$ individuals is then
    \begin{equation}
        q_n(a) = \mathbb{P}\bigl(N(a)=0 \mid N(\mA)=n\bigr)
        =\left(1-\frac{a}{|\mA|}\right)^{n} = q_1^n,
    \end{equation}
    Therefore, we have
    \begin{equation}
    	p_0(a)
    	=G\{f^+_n\}(1-a/|\mA|),
    	\label{eq:PGF_cond}
    \end{equation}
    where $G\{f^+_{n,\vartheta}\}(q_1):=\sum_{n\ge1}f^+_{n,\vartheta}q_1^n$ is the probability-generating function (PGF) of the SAD with parameters $\vartheta$. 
    Species richness versus effort
    as union of the sampled area is then
    \begin{equation}
        \cS_\vartheta(a) =S_{\mathrm{T}}\bigl[1-G\{f^+_n\}(1-a/|\mA|)\bigr],
    \end{equation}
    \citep{Coleman1981,he2002}.
    Any SAD with a closed form PGF will thus also have a closed form richness-effort curve.
    For instance, when $f^+_{n,\vartheta}$ is a geometric distribution with parameter $\vartheta$,
    then the richness-effort curve is the classic Michaelis--Menten type functional form $\cS_\vartheta(\alpha)=S_\mathrm{T}\tfrac{\alpha}{\vartheta+(1-\vartheta)\alpha}$, with $    \alpha:=a/|\mA|$ the fraction of the unit covered by samples.
    
    \paragraph{Richness versus sum of sampled areas.} We instead quantify
    sampling effort as the sum of sampled areas: 
    \begin{equation}
        a_{\is} := \sum_{i\in\is} |\ma_i|.
    \end{equation}
    The possibility of overlap means that a specific individual can be ``captured'' multiple times. Under URP and the sampling process described above, 
    with $|\ma_i|\ll|\mA|$, the number of captures of a given individual is Poisson with intensity $\lambda = a/|\mA|\in[0,\infty)$, where uniformity permits dropping the $\is$. Thus, for a species with $n$ individuals, the total number of captures $C(a)$ follows:
    \begin{equation}
        C(a)\mid N(\mA)=n \sim \mathrm{Pois}\!\left(n\frac{a}{|\mA|}\right),
    \end{equation}
    so the probability of missing a species with $n$ individuals is
    \begin{equation}
        q_n(a) = \mathbb{P}\bigl(C(a)=0 \mid N(\mA)=n\bigr)
        = \exp\left(-n\frac{a}{|\mA|}\right)=q_1^n,
    \end{equation}
    Therefore, we have
    \begin{equation}
    	p_0(a)
    	=G\{f^+_n\}(e^{-a/|\mA|}).
    \end{equation}
    such that
    \begin{equation}
        \cS_\vartheta(a) =S_{\mathrm{T}}\bigl[1-G\{f^+_n\}(e^{-a/|\mA|})\bigr].
    \end{equation}
    Again, any SAD with a closed form PGF will thus also have a closed form richness-effort curve.
    For instance,
    when $f^+_{n,\vartheta}$ is a geometric distribution with parameter $\vartheta$,
    then the rarefaction curve is $\cS_\vartheta(\beta)=S_\mathrm{T}\tfrac{1-e^{-\beta}}{1-(1-\vartheta)e^{-\beta}}$, with $\beta:=a/|\mA|$, where $a$ is the sum of sample areas.
    
    \newpage
    \begin{landscape}
    \begin{table}[p]
    	\section{Supplementary tables}
    	\centering
    	\small
    	\setlength{\tabcolsep}{6pt}
    	\renewcommand{\arraystretch}{1.2}
    	\begin{tabular}{p{1.8cm} p{5.6cm} p{4.1cm} p{5.6cm} p{5.6cm}}
    		\hline
    		\textbf{SAD} (ZT)
    		& \(\boldsymbol{f^+_n}\) (PMF over species)
    		& \(\boldsymbol{G\{f^+_{n,\vartheta}\}(z)}\)
    		& \(\boldsymbol{\cS_\vartheta(\alpha)=S_{\mathrm T}[1-G\{f^+_{n,\vartheta}\}(1-\alpha)]}\)
    		& \(\boldsymbol{\cS_\vartheta(\beta)=S_{\mathrm T}[1-G\{f^+_{n,\vartheta}\}(e^{-\beta})]}\)\\
    		\hline
    
    		$\mathrm{Logseries}(x)$ &
    		\(\dfrac{x^n}{n\,[-\ln(1-x)]},\; n\ge1\) &
    		\(\dfrac{\ln(1-xz)}{\ln(1-x)}\) &
    		\(\displaystyle S_{\mathrm T}\!\biggl[1-\dfrac{\ln\!\bigl(1-x(1-\alpha)\bigr)}{\ln(1-x)}\biggr]\) &
    		\(\displaystyle S_{\mathrm T}\!\biggl[1-\dfrac{\ln\!\bigl(1-xe^{-\beta}\bigr)}{\ln(1-x)}\biggr]\) \\[.8ex]
    
    		\(\mathrm{Pois}^{+}(\mu)\) &
    		\(\dfrac{e^{-\mu}\mu^{n}/n!}{1-e^{-\mu}},\; n\ge1\) &
    		\(\dfrac{e^{\mu(z-1)}-e^{-\mu}}{1-e^{-\mu}}\) &
    		\(\displaystyle S_{\mathrm T}\,\dfrac{1-e^{-\mu\alpha}}{1-e^{-\mu}}\) &
    		\(\displaystyle S_{\mathrm T}\,\dfrac{1-\exp\!\bigl[-\mu(1-e^{-\beta})\bigr]}{1-e^{-\mu}}\) \\[.8ex]
    
    		\(\mathrm{NB}^{+}(r,p)\) &
    		\(\dfrac{\binom{n+r-1}{n}(1-p)^n p^{r}}{1-p^{r}},\; n\ge1\) &
    		\(\dfrac{\left(\dfrac{p}{1-(1-p)z}\right)^{r}-p^{r}}{1-p^{r}}\) &
    		\(\displaystyle S_{\mathrm T}\,
    		\dfrac{1-\left(\dfrac{p}{\,p+(1-p)\alpha\,}\right)^{r}}{1-p^{r}}\) &
    		\(\displaystyle S_{\mathrm T}\,
    		\dfrac{1-\left(\dfrac{p}{\,1-(1-p)e^{-\beta}\,}\right)^{r}}{1-p^{r}}\) \\[.8ex]
    
    		\(\mathrm{Geom}^{+}(p)^*\) &
    		\(p(1-p)^{\,n-1},\; n\ge1\) &
    		\(\dfrac{p\,z}{1-(1-p)z}\) &
    		\(\displaystyle S_{\mathrm T}\,\dfrac{\alpha}{\,p+(1-p)\alpha\,}\) &
    		\(\displaystyle S_{\mathrm T}\,\dfrac{1-e^{-\beta}}{\,1-(1-p)e^{-\beta}\,}\) \\[.8ex]
    
    		Most even &
    		\(f^+_{n^\ast}=1\ \text{with }n^\ast=N/S_{\mathrm T}\in\mathbb{N};\ \text{rest }0\) &
    		\(z^{\,N/S_{\mathrm T}}\) &
    		\(\displaystyle S_{\mathrm T}\bigl[1-(1-\alpha)^{N/S_{\mathrm T}}\bigr]\) &
    		\(\displaystyle S_{\mathrm T}\bigl[1-e^{-\beta\,N/S_{\mathrm T}}\bigr]\) \\[.8ex]
    
    		Most uneven &
    		\(f^+_1=\dfrac{S_{\mathrm T}-1}{S_{\mathrm T}},\ 
    		f^+_{\,N-S_{\mathrm T}+1}=\dfrac{1}{S_{\mathrm T}};\ \text{rest }0\) &
    		\(\dfrac{S_{\mathrm T}-1}{S_{\mathrm T}}\,z+\dfrac{1}{S_{\mathrm T}}\,z^{\,N-S_{\mathrm T}+1}\) &
    		\(\displaystyle 1+(S_{\mathrm T}-1)\alpha-(1-\alpha)^{\,N-S_{\mathrm T}+1}\) &
    		\(\displaystyle 1+(S_{\mathrm T}-1)\bigl(1-e^{-\beta}\bigr)-e^{-\beta\,(N-S_{\mathrm T}+1)}\) \\[.8ex]
    
    		\hline
    	\end{tabular}
    	\caption{\textbf{Expected species richness under uniform random placement (URP) for a selection of zero-truncated species abundance distributions (SADs) and different sampling assumptions}. Union-based effort $\alpha$ is sampled area when sampling in disjoint subsets. The expressions in terms of union-based effort were derived before in e.g. \citealp{Coleman1981,he2002}. Sum-based effort $\beta$, what we used, is sampled area when allowing many small biodiversity plots to overlap. Both effort types are in units of area relative
        to $|\mA|$ here. \(^{*}\)\(\mathrm{Geom}^{+}(p)=\mathrm{NB}^{+}(1,p)\). See \cref{figSI:rarefactionURP} for the corresponding plots.}
    	\label{tab:ESA_binom_SADs}
    \end{table}
    \end{landscape}

    \begin{table}[h!]
        \centering
        \small
        \setlength{\tabcolsep}{4pt}
        \begin{tabularx}{\textwidth}{l l >{\centering\arraybackslash}X >{\centering\arraybackslash}X >{\centering\arraybackslash}X >{\centering\arraybackslash}X >{\centering\arraybackslash}X}
        \toprule
        Model & Predictors & RMSE & MAPE & Rel. Bias & $R^2$ & $D^2$ \\
        \midrule
        FFNN & Env. + Area & 47.431 ± 2.911 & \textbf{0.564 ± 0.010} & 0.041 ± 0.013 & \textbf{0.988 ± 0.002} & \textbf{0.855 ± 0.008} \\
        MuScaRi & Area & 161.048 ± 5.309 & 0.989 ± 0.269 & 0.227 ± 0.142 & 0.857 ± 0.006 & 0.601 ± 0.023 \\
        MuScaRi & Env. & 48.242 ± 1.535 & 0.582 ± 0.034 & \textbf{0.007 ± 0.044} & 0.987 ± 0.001 & 0.848 ± 0.009 \\
        MuScaRi & Env. + Area & \textbf{47.206 ± 1.252} & 0.585 ± 0.046 & 0.025 ± 0.048 & \textbf{0.988 ± 0.001} & 0.851 ± 0.008 \\
        \bottomrule
        \end{tabularx}
        \caption{\textbf{Interpolation performance on the EVA test dataset (mean ± standard deviation across split).} RMSE: root mean squared error; MAPE: mean absolute percentage error; Rel.\ Bias: median relative bias, defined as $\mathrm{median}[(\hat{S}-S)/S]$; $R^2$: coefficient of determination; $D^2$: fraction of deviance explained. Lower RMSE, MAPE, and Rel.\ Bias (closer to 0), and higher $R^2$ and $D^2$, indicate better performance.}
        \label{tab:interp_performance}
    \end{table}

    \FloatBarrier

    \begin{table}[h!]
        \centering
        \small
        \setlength{\tabcolsep}{4pt}
        \begin{tabularx}{\textwidth}{l l >{\centering\arraybackslash}X >{\centering\arraybackslash}X >{\centering\arraybackslash}X >{\centering\arraybackslash}X >{\centering\arraybackslash}X}
        \toprule
        Model & Predictors & RMSE & MAPE & Rel. Bias & $R^2$ & $D^2$ \\
        \midrule
        FFNN & Env. + Area & 2054.773 ± 571.212 & 1.708 ± 0.475 & 1.241 ± 0.349 & -4.848 ± 2.778 & -1.613 ± 0.789 \\
        Chao2 estimator & -- & 1258.533 ± 11.719 & 0.610 ± 0.010 & -0.630 ± 0.017 & -1.174 ± 0.196 & -0.654 ± 0.096 \\
        MuScaRi & Area & 943.764 ± 337.362 & 0.724 ± 0.397 & 0.111 ± 0.498 & -0.292 ± 1.034 & -0.105 ± 0.475 \\
        MuScaRi & Env. & 623.824 ± 97.925 & 0.651 ± 0.117 & 0.283 ± 0.088 & 0.487 ± 0.163 & 0.270 ± 0.136 \\
        MuScaRi & Env. + Area & \textbf{498.519 ± 32.914} & \textbf{0.378 ± 0.071} & \textbf{-0.066 ± 0.060} & \textbf{0.679 ± 0.042} & \textbf{0.430 ± 0.063} \\
        \bottomrule
        \end{tabularx}
        \caption{\textbf{Extrapolation performance on the GIFT dataset under asymptotic sampling effort (mean ± standard deviation across splits).} RMSE: root mean squared error; MAPE: mean absolute percentage error; Rel.\ Bias: median relative bias, defined as $\mathrm{median}[(\hat{S}-S)/S]$; $R^2$: coefficient of determination; $D^2$: fraction of deviance explained. Lower RMSE, MAPE, and Rel.\ Bias (closer to 0), and higher $R^2$ and $D^2$, indicate better performance.}
        \label{tab:extrap_performance}
    \end{table}
    \FloatBarrier

    \begin{table}[h!]
        \centering
        \small
        \setlength{\tabcolsep}{6pt}
        \begin{tabularx}{\textwidth}{l >{\centering\arraybackslash}X >{\centering\arraybackslash}X >{\centering\arraybackslash}X}
        \toprule
        Split & Train & Validation & Test \\
        \midrule
        0 & 216400 & 65444 & 71064 \\
        1 & 211624 & 75840 & 65444 \\
        2 & 204272 & 72796 & 75840 \\
        3 & 212348 & 67764 & 72796 \\
        4 & 214080 & 71064 & 67764 \\
        \bottomrule
        \end{tabularx}
        \caption{\textbf{Sample counts per training, validation, and test datasets for each training/validation/test split.}}
        \label{tab:fold_sample_counts}
    \end{table}
    \FloatBarrier

    \begin{table}[h!]
        \centering
        \small
        \setlength{\tabcolsep}{5pt}
        \begin{tabularx}{\textwidth}{l r r >{\centering\arraybackslash}X >{\centering\arraybackslash}X >{\centering\arraybackslash}X >{\centering\arraybackslash}X}
        \toprule
        Location & Lat & Lon & $S_T$ at 25km$^2$ & $S_T$ at 2500km$^2$ & $\frac{d \log(S_T)}{d \log(A)}$ at 25km$^2$ &  $\frac{d \log(S_T)}{d \log(A)}$ at 2500km$^2$ \\
        \midrule
        A & 49.66 & 16.00 & 900.596 ± 231.857 & 1314.927 ± 101.480 & 0.069 ± 0.048 & 0.045 ± 0.029 \\
        B & 46.67 & 10.20 & 1028.228 ± 228.334 & 1507.660 ± 161.564 & 0.095 ± 0.091 & 0.093 ± 0.079 \\
        C & 52.05 & 6.02 & 868.746 ± 244.337 & 1102.179 ± 165.779 & 0.331 ± 0.145 & 0.031 ± 0.030 \\
        \bottomrule
        \end{tabularx}
        \caption{\textbf{Pointwise estimate of species richness and rate of species accumulation obtained for nested spatial units with sizes 25km$^2$ and 2500km$^2$ and different locations, as reported in \cref{fig:maps}.} Values are mean ± standard deviation across ensemble members.}
        \label{tab:sar_summary}
    \end{table}

    \pagebreak

    \section{Supplementary figures} \label{secSI:sup_mat}

    \begin{figure}[h!] 
        \centering
        \includegraphics[]{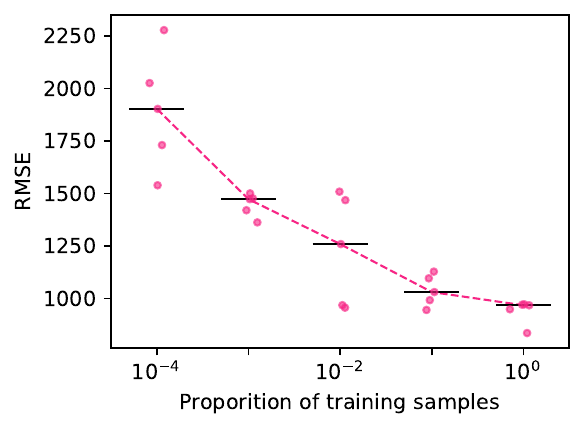}
        \caption{\textbf{Model's performance in predicting the EVA-based test splits against the proportion of generated training samples relative to the number of vegetation plots}.} \label{figSI:scaling_exp}
    \end{figure}
    \FloatBarrier

    \begin{figure}[h!] 
        \centering
        \includegraphics[width=0.9\textwidth]{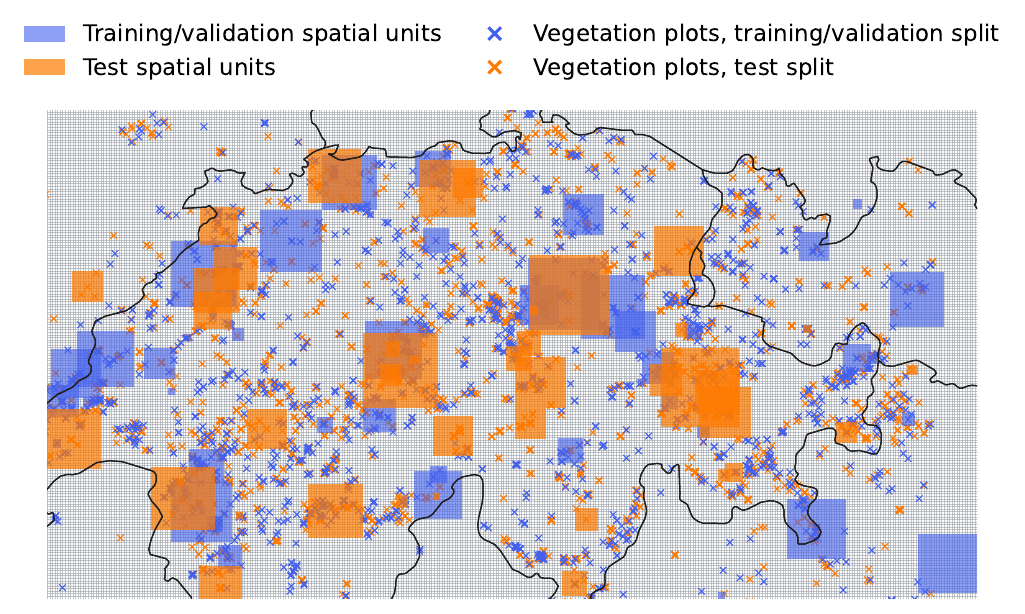}
        \caption{\textbf{Spatial blocking procedure for generating the training, validation and test splits}. 
        EVA plots (crosses) are shown for a close-up of Switzerland, overlaid on the 1\,km $\times$ 1\,km spatial blocks used to assign plots to five distinct folds. Blue crosses belong to the training/validation folds; orange crosses belong to the test fold for this specific split. Training, validation, and test samples (each sample corresponding to a triplet of sampling effort, spatial-unit features, and observed species richness) are generated independently from the plots of their respective assigned fold (see \cref{sec:data}). This ensures each plot appears in exactly one of the five validation and test split, and prevents data leakage even when spatial units from the training and validation or test datasets geographically overlap. Only a subset of plots and spatial units is shown for clarity.} \label{figSI:spatial_blocks}
    \end{figure}
    \FloatBarrier

    \begin{figure}[h!] 
        \centering
        \includegraphics[width=0.6\textwidth]{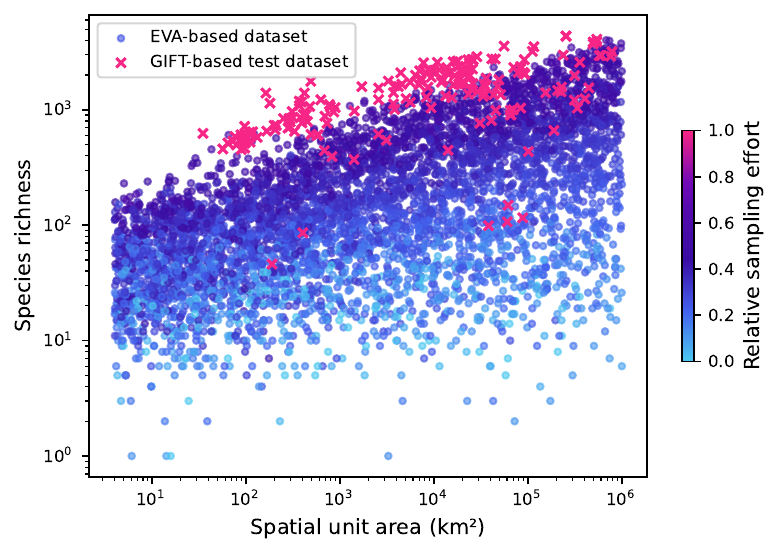}
        \caption{\textbf{Species richness against spatial unit area and relative sampling effort, for the one of the EVA-based cross-validation split and the GIFT-based test dataset}. The relative sampling effort is calculated as the ratio of the logarithm of sampling effort $a$ over the logarithm of the spatial unit area.} \label{figSI:datasets}
    \end{figure}
    \FloatBarrier

    \begin{figure}[h!] 
        \centering
        \includegraphics[width=0.9\textwidth]{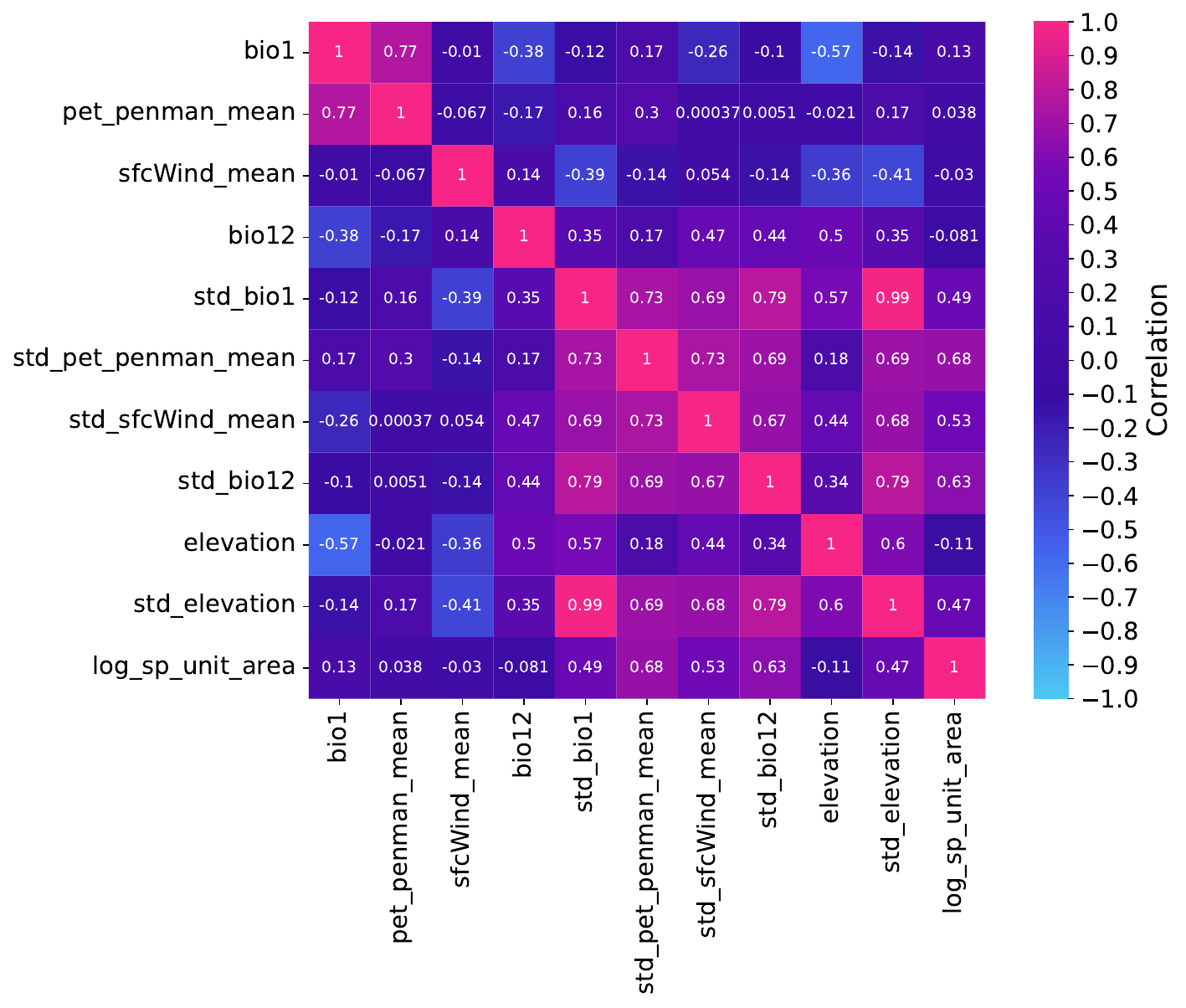}
        \caption{\textbf{Correlation matrix of features used for model training and inference}.} \label{figSI:corr_mat}
    \end{figure}
    \FloatBarrier




    \FloatBarrier

    \begin{figure}[h!] 
        \centering
        \includegraphics[width=\textwidth]{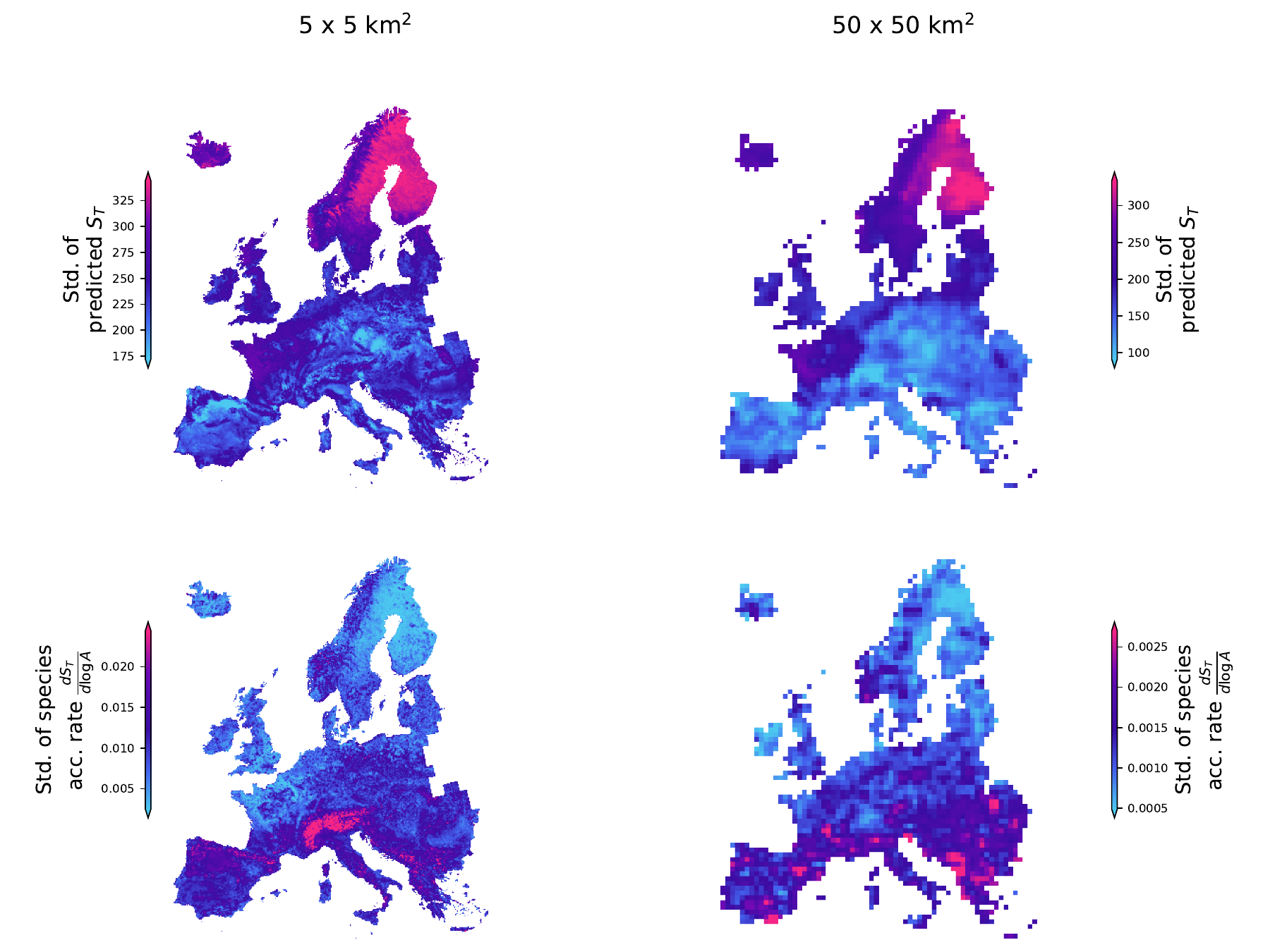}
        \caption{\textbf{Standard deviation of the predictions obtained from the \modelname{} ensemble model and displayed in \cref{fig:maps}}.} \label{figSI:maps_std}
    \end{figure}
    \FloatBarrier

    \begin{figure}[h!] 
        \centering
        \includegraphics[width=0.9\textwidth]{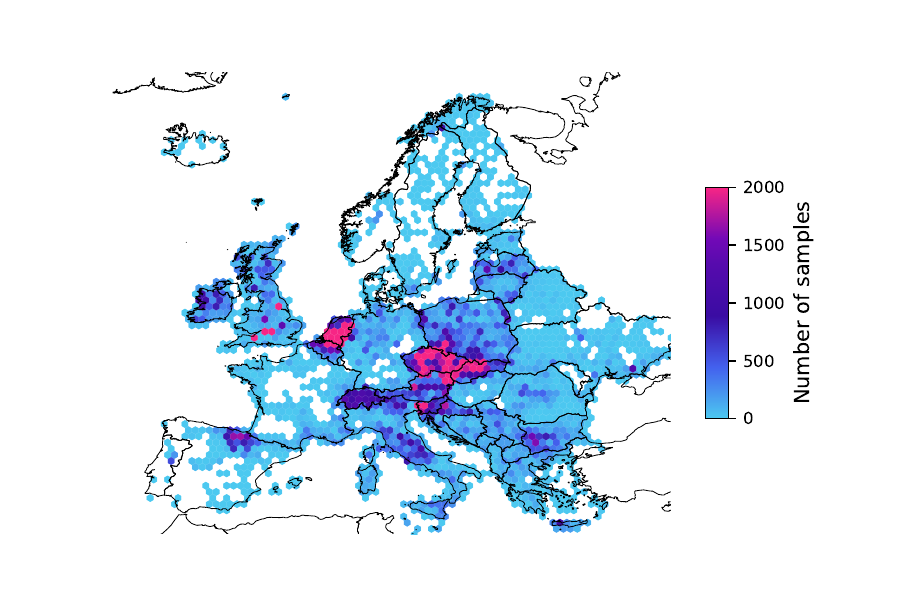}
        \caption{\textbf{Density of EVA vegetation plots (number of samples per hexagon with width of 50km)}.} \label{figSI:EVA_locations}
    \end{figure}
    \FloatBarrier

\end{document}